\documentclass[lettersize,journal]{IEEEtran}
\usepackage{amsmath,amsfonts}
\usepackage{algorithmic}
\usepackage{textcomp}
\usepackage{xcolor}
\usepackage{booktabs}
\usepackage[colorlinks,linkcolor=blue]{hyperref}
\usepackage{array}
\usepackage{multirow}
\usepackage{multicol}
\usepackage{enumitem}
\usepackage{graphicx}
\usepackage{subfigure}
\usepackage{bm}
\usepackage{tabularx}
 
\usepackage{amssymb}
\usepackage[normalem]{ulem}
\useunder{\uline}{\ul}{}

\newcommand{\MLP}{\mathrm{MLP}}

\begin{document}
\title{Pattern-wise Transparent Sequential Recommendation}
\author{
        Kun~Ma,
        Cong~Xu,
        Zeyuan~Chen,
        Wei~Zhang,~\IEEEmembership{Member,~IEEE}
\thanks{
\IEEEcompsocthanksitem 
This work was supported in part by the National Natural Science Foundation of China under Grant 92270119, Grant 62072182, and the Key Laboratory of Advanced Theory and Application in Statistics and Data Science, Ministry of Education. (Corresponding author: Wei Zhang).
}
\thanks{
\IEEEcompsocthanksitem 
	Kun Ma, Cong Xu, and Wei Zhang are with the School of Computer Science and Technology,
	East China Normal University, Shanghai, China. Wei Zhang is also with Shanghai Innovation Institute (email: 51255901097@stu.ecnu.edu.cn, congxueric@gmail.com, zhangwei.thu2011@gmail.com). 
}
\thanks{
\IEEEcompsocthanksitem 
        Zeyuan Chen is with the Ant Group, Hangzhou, China.
        (email: chenzeyuan.czy@antgroup.com)
}
\thanks{Manuscript received XX XX, 2023; revised XX XX, XXXX.}
}%

\markboth{Journal of \LaTeX\ Class Files,~Vol.~14, No.~8, August~2021}%
{Shell \MakeLowercase{\textit{et al.}}: A Sample Article Using IEEEtran.cls for IEEE Journals}

\maketitle

\begin{abstract}
A transparent decision-making process is essential for developing reliable and trustworthy recommender systems. 
For sequential recommendation, it means that the model can identify key items that account for its recommendation results. 
However, achieving both interpretability and recommendation performance simultaneously is challenging, especially for models that take the entire sequence of items as input without screening.
In this paper, we propose an interpretable framework (named PTSR) 
that enables a pattern-wise transparent decision-making process without extra features. 
It breaks the sequence of items into multi-level patterns 
that serve as atomic units throughout the recommendation process.
The contribution of each pattern to the outcome is quantified in the probability space. 
With a carefully designed score correction mechanism,
the pattern contribution can be implicitly learned in the absence of ground-truth key patterns.
The final recommended items are those that most key patterns strongly endorse.
Extensive experiments on five public datasets demonstrate remarkable recommendation performance, 
while statistical analysis and case studies validate the model interpretability.
\end{abstract}

\begin{IEEEkeywords}
Recommender system, sequential recommendation, interpretability, transparent model
\end{IEEEkeywords}

\section{Introduction}\label{sec:introduction}

\begin{figure}[!t]
    \centering
    \includegraphics[width=\linewidth]{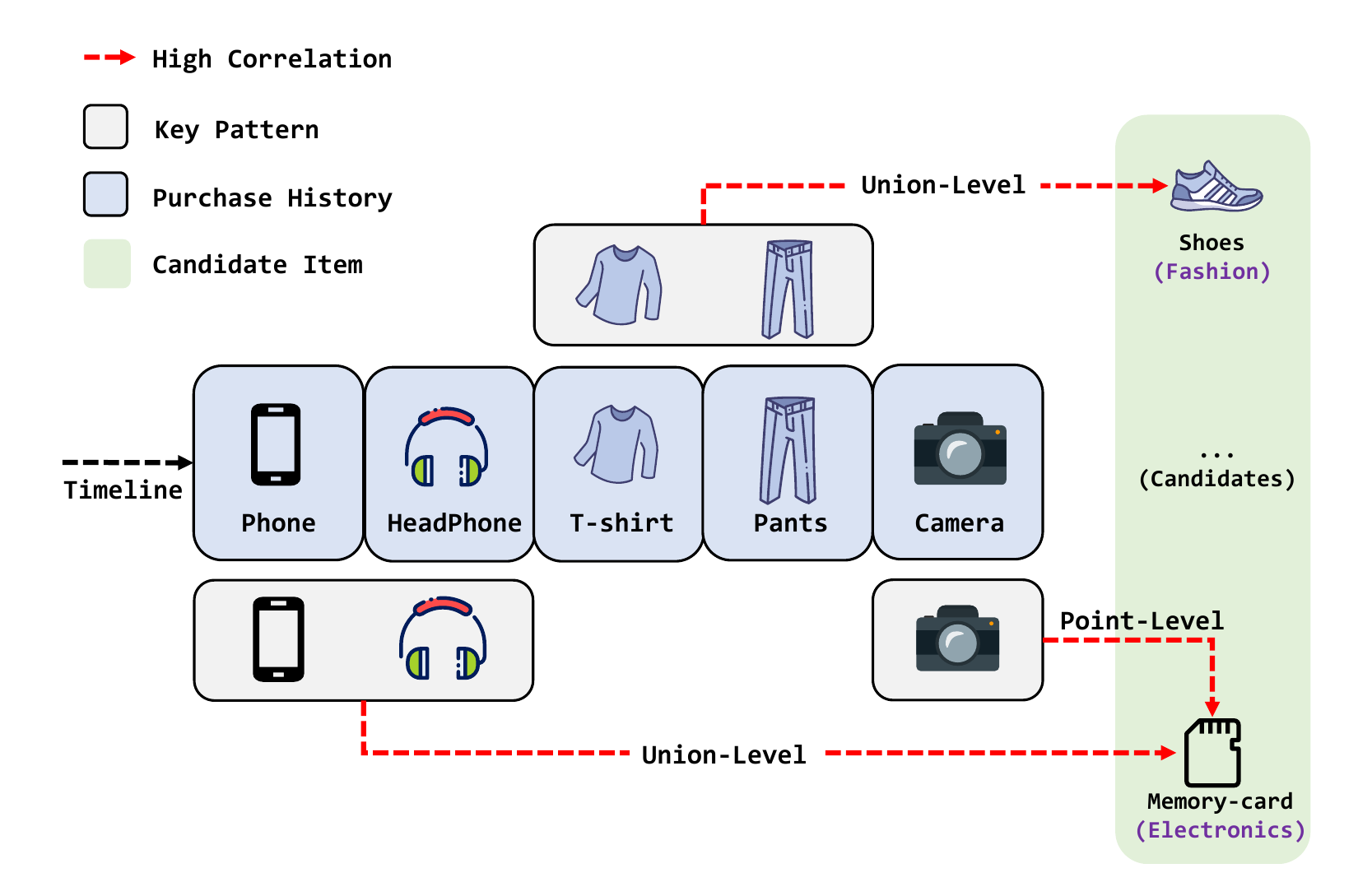}
    \caption{
    Illustration of interpretability at the point level and union level.
    The former is commonly used in existing studies to identify key individual items, 
    while the latter provides additional insights into the common interests exhibited by a user.
    }
    \label{fig:multi-level-case}
\end{figure}

\IEEEPARstart{R}{ecommender} systems have been engineered 
to expedite the process of identifying items 
that align with users' interests and to do so with a high degree of precision. 
In recent years, how to design decision-transparent models 
to improve the interpretability of recommender systems has become a popular topic~\cite{DBLP:journals/tkde/HeHSLJC18/NAIS, DBLP:conf/wsdm/ChenXZT0QZ18/RUM, DBLP:journals/ftir/ZhangC20/ExplainRec}. 
Different from traditional approaches, in addition to providing highly accurate candidate items,
a good interpretable model needs to have the ability to provide compelling and easy-to-understand explanations for its recommendations \cite{DBLP:journals/ftir/ZhangC20/ExplainRec}
--- how the input (i.e., interacted items) affects the output (i.e., candidate items to be recommended).
In the context of ID-based sequential recommendation scenarios, 
explanations typically refer to the distribution of item importance within interaction sequences. 
In general, the explanations should \textbf{make sense} to people,  
and they should also be \textbf{consistent} with the model's decisions~\cite{DBLP:conf/acl/SerranoS19/AttentionNotExp}.  
This not only helps to increase user confidence by providing explanations 
but also helps engineers diagnose the causes of bad cases.
Despite extensive exploration in this field, two main challenges remain to be addressed:
\begin{itemize}[leftmargin=*]
  \item \textbf{How to simultaneously achieve high performance and transparency.}
  Existing interpretable models, exemplified by NAIS~\cite{DBLP:journals/tkde/HeHSLJC18/NAIS} and RUM~\cite{DBLP:conf/wsdm/ChenXZT0QZ18/RUM}, 
  achieve transparency through explicit relevance computation between target items and historical sequences. 
  While ensuring interpretability, 
  it often results in suboptimal performance (see Table~\ref{tbl:performance-cmp}), 
  primarily attributable to their inability to capture high-order item interactions.
  Furthermore, these models do not incorporate additional designs to compensate for the performance degradation caused by their simplistic architectures.
  In contrast, high-performance models often fall short in transparency, 
  as they usually involve complex nonlinear transformations to enable more comprehensive embedding learning. 
  For instance, attention-based methods~\cite{DBLP:conf/icdm/KangM18/SASRec, DBLP:conf/cikm/SunLWPLOJ19/Bert4Rec} utilize attention weights as a means of interpretability, 
  but studies~\cite{BaiLZLBW21,LopardoPG24} have questioned the consistency between these weights and the model's decision-making, 
  casting doubt on their interpretability. 
  Consequently, designing models that successfully balance performance and transparency remains an area for further exploration.

  \item \textbf{How to move beyond point-level interpretability to union-level interpretability.}
  Some studies~\cite{0ddadb0a-e924-3251-a5fc-c143b9c1b422/PIBP, DBLP:conf/sigir/ZhangPL0L24/DPNCTR} suggest that 
  both point-level (i.e., individual items) and union-level (i.e., subsequences of items) patterns are correlated with user personality. 
  Therefore, it is essential to provide explanations that encompass both levels of patterns.
  As shown in Fig.~\ref{fig:multi-level-case}, 
  the rating for the `Memory-card' is directly influenced by the `Camera', 
  while the electronic pattern composed of the `Phone' and `Headphone' also plays a significant role that cannot be overlooked.
  Existing methods~\cite{DBLP:journals/tkde/HeHSLJC18/NAIS, DBLP:conf/wsdm/ChenXZT0QZ18/RUM} based on target attention are confined to point-level interpretability; 
  other approaches capable of modeling union-level patterns also exhibit limitations. 
  For instance, some methods~\cite{DBLP:conf/wsdm/TangW18/Caser, DBLP:conf/wsdm/Guo0SZWBZ22/MSGIFSR} employ pooling operations to fuse embeddings, modeling patterns as fixed points in vector space, 
  which may fail to capture the combinatorial effects among items~\cite{DBLP:conf/www/FanLWWNZPY22/STOSA, DBLP:conf/nips/RenL20/BetaE}.
  Furthermore, alternative methods~\cite{DBLP:conf/wsdm/Guo0SZWBZ22/MSGIFSR, DBLP:conf/kdd/Wang00LC19/KGAT} utilize graph networks to model complex relationships between items, 
  but the downside is the obscurity of the components of high-order patterns 
  and the ambiguity of each high-order pattern's contribution, 
  resulting in an inability to manifest union-level interpretability.
\end{itemize}

Simultaneously addressing both challenges is non-trivial. 
Firstly, for black-box models, increasing the depth of network to enhance model's nonlinear fitting capacity generally contributes to improved performance~\cite{DBLP:conf/cvpr/HeZRS16/ResNet, DBLP:conf/ijcai/LiZL0WG22/MLP4Rec}. 
However, this comes at the cost of reduced transparency. 
Therefore, further exploration is required to enhance performance under the condition of a simple architecture.
Secondly, union-level interpretability inherently requires decomposable pattern structures, 
necessitating explicit pattern extraction before modeling intra-pattern item relationships~\cite{DBLP:conf/wsdm/Guo0SZWBZ22/MSGIFSR, DBLP:conf/sigir/ZhangPL0L24/DPNCTR}. 
This constraint prohibits direct embedding fusion as employed in Transformer-based methods~\cite{DBLP:conf/icdm/KangM18/SASRec, DBLP:conf/cikm/SunLWPLOJ19/Bert4Rec, DBLP:conf/wsdm/LiWM20/TiSASRec}, 
where the interpretability of attention weights in deep layers remains questionable~\cite{DBLP:conf/acl/SerranoS19/AttentionNotExp, DBLP:conf/naacl/JainW19/AttentionNotExp2}. 
Thirdly, union-level interpretability must demonstrate dual benefits: enriching interpretability while maintaining or improving performance. 
Despite existing evidence~\cite{0ddadb0a-e924-3251-a5fc-c143b9c1b422/PIBP, DBLP:conf/sigir/ZhangPL0L24/DPNCTR} linking user personality traits to union-level patterns, 
real-world scenarios lack ground-truth for patterns. 
If the model incorrectly combines unrelated items into patterns, it can lead to performance degradation.

Taking into account these issues, we propose PTSR, a novel \underline{P}attern-wise \underline{T}ransparent \underline{S}equential \underline{R}ecommendation model.
Its simple architecture ensures transparency in the decision-making process, 
while the introduction of probabilistic embeddings enhances the accuracy of pattern modeling. 
Coupled with a weighting module, these components collectively ensure the model's performance.

Specifically, our model employs target attention mechanism that 
directly computes relevance between target item and patterns.
To enable this transparent architecture to surpass the performance of black-box models, 
we focus on enhancing the accuracy of pattern representation modeling:
(1) We utilize sliding windows of varying sizes to extract point-level and union-level patterns explicitly. This ensures the decomposability of pattern components;
(2) Typical embeddings is replaced with probabilistic embeddings to refine representation granularity and model uncertainty simultaneously. 
Studies~\cite{DBLP:journals/vc/KarpukhinDK24/PER, DBLP:conf/iccv/ShiJ19/PFE, DBLP:journals/corr/abs-1909-11702/SPE} have shown that representing items as fixed points in space is inadequate for capturing the nuances of differences between items and contradicts the inherent uncertainty~\cite{DBLP:conf/www/FanLWWNZPY22/STOSA, DBLP:conf/ijcai/YuanZX0LS023/SRPLR} in recommendation systems.
(3) Within patterns, unlike methods~\cite{DBLP:conf/wsdm/TangW18/Caser, DBLP:conf/wsdm/Guo0SZWBZ22/MSGIFSR} that use pooling or GRU for representation aggregation, 
we propose employing conjunction operator to capture item commonalities from logical perspective, 
thereby better leveraging the model's cognitive capabilities~\cite{DBLP:conf/www/ChenSLZ21/NCR, DBLP:conf/cikm/ShiCMMZZ20/NLR}.
After obtaining pattern representations, we use KL-Divergence to calculate their `distance' from target item.

During the training process, 
we observe that the absence of ground-truth for patterns prevents the model from highlighting effective patterns when directly optimizing the distance, leading to suboptimal performance. 
To address this, we design a weighting strategy that adaptively enhances the optimization of key patterns with smaller distances by transforming negative distances into weights, 
effectively improving the model's performance. Additionally, unlike traditional positional encoding, 
we introduce a sequence bias to enable the model to adjust recommendations when the sequence order changes.

In general, the contribution of this paper can be summarized as follows:
\begin{itemize}[leftmargin=*]
    \item
    We propose PTSR, a novel model that simultaneously enhances recommendation performance and interpretability through multi-level pattern learning.
    
    \item
    We introduce probabilistic embeddings to enhance representation granularity and model uncertainty, while employing logical operators for item aggregation.
    
    \item
    We design a distance-based weight component for pattern optimization and sequence-aware bias for positional encoding.
    
    \item
    Extensive experiments on 5 datasets demonstrate PTSR's superior performance over 11 baselines, with case studies and statistical analysis validating its interpretability.
\end{itemize}

\section{Related Work}\label{sec:related-work}
\subsection{Sequential Recommendation}
Recommender systems are designed to retrieve items of interest to the user. 
Some early studies, such as general collaborative filtering
~\cite{DBLP:conf/www/HeLZNHC17/NCF, DBLP:conf/kdd/Koren09/CFTD}, adopt simple architectures,
thereby achieving decent efficiency and a certain degree of interpretability (mostly point-level).
Despite their satisfactory efficiency, the lack of sequential information confines them to static user profiles.
Therefore, sequential models~\cite{DBLP:conf/wsdm/LiWM20/TiSASRec} are gaining increasing popularity within the research community due to their potential in dynamic interest modeling.
The development of this field has closely followed the evolution of sequential modeling:
embarking from traditional models \cite{DBLP:conf/www/RendleFS10/FPMC,DBLP:conf/icdm/HeM16/Fossil} based on Markov chains, 
to more recent techniques such as GRU4Rec \cite{DBLP:journals/corr/HidasiKBT15/GRU4Rec,DBLP:conf/cikm/HidasiK18/GRU4Rec+} and SASRec \cite{DBLP:conf/icdm/KangM18/SASRec},
where recurrent neural networks \cite{DBLP:conf/ssst/ChoMBB14/GRU} and transformers \cite{DBLP:conf/nips/VaswaniSPUJGKP17/Transformer} are employed, respectively.
These sequential models, particularly attention-based approaches \cite{DBLP:conf/icdm/KangM18/SASRec,DBLP:conf/wsdm/LiWM20/TiSASRec,DBLP:conf/cikm/SunLWPLOJ19/Bert4Rec},
have demonstrated impressive recommendation performance.

However, the increasing precision comes with rapidly deteriorating model transparency. 
Compared to general collaborative filtering, both the gated hidden unit utilized in GRU4Rec and the autoregressive attention mechanism employed in SASRec are more sophisticated yet lack interpretability.
Consequently, they often fail to identify key items to support their predictions \cite{DBLP:conf/acl/SerranoS19/AttentionNotExp,DBLP:conf/naacl/JainW19/AttentionNotExp2,DBLP:conf/emnlp/WiegreffeP19/AttentionNotNotExp}.
Although some efforts have been made to simplify the architectures (e.g., FMLP-Rec~\cite{DBLP:conf/www/ZhouYZW22/FMLPRec} is a pure MLP model), 
the decision transparency is inevitably compromised as more blocks are stacked to achieve sufficient model expressive power.
In contrast, PTSR proposed in this paper is decision-transparent at the model architecture level while preserving satisfactory recommendation performance.

\begin{table}[]
\centering
\renewcommand\arraystretch{1.2}
\caption{Notations used in this paper}
\label{tbl:notation}
\scalebox{1.0}{
    \begin{tabularx}{\linewidth}{l@{\hspace{1.5cm}}l}
    \toprule
    Notation & Description \\ \hline
    $s$      & user interaction sequence \\
    $v$      & item \\
    $l$      & level of pattern \\ 
    $p^{(l)}_k$  & the $k$-th pattern of the $l$-th level \\
    $\alpha$, $\beta$  & parameters of the Gamma distribution \\
    $f(\cdot;\alpha, \beta)$  & probability density function (PDF) \\
    $\eta$  & distance-based weight \\
    $\delta$  & sequence-aware bias \\
    $\gamma$  & margin parameter \\
    $\mathbf{v} \in \mathbb{R}_{+}^{2d}$  & item representation \\
    $\mathbf{p} \in \mathbb{R}_{+}^{2d}$  & pattern representation \\
    \bottomrule
    \end{tabularx}
}
\end{table}

\subsection{Explainable Recommendation}
Compared to black-box models, 
explainable recommender systems enhance user trust by offering rationale for their recommendations, while maintaining high performance. 
In recent years, numerous significant studies~\cite{DBLP:journals/tkde/HeHSLJC18/NAIS, DBLP:conf/wsdm/ChenXZT0QZ18/RUM, DBLP:conf/www/ZhangYWW22/NS-ICF, DBLP:conf/sigir/ZhangL0ZLM14/EFM, DBLP:conf/cikm/TanXG00Z21/CountER} have emerged in this field. 
These can be broadly categorized into two main groups based on their use of side information: 
similarity-based and aspect-based.

Similarity-based approaches usually take only the itemID as input 
and learn the relationship between items based on a large amount of interaction data. 
Its interpretability is demonstrated by showing the weights of the items that users have interacted with.
For example, Abdollahi et al.~\cite{DBLP:conf/icml/AbdollahiN16/Beh} introduces the use of Restricted Boltzmann Machines (RBM) 
to generate explainable top-N recommendation lists. 
Similarly, EMF~\cite{DBLP:conf/recsys/AbdollahiN17/Beh2} employs 
constrained matrix factorization for interpretable recommendations, 
mitigating the trade-off between interpretability and accuracy. 
NAIS~\cite{DBLP:journals/tkde/HeHSLJC18/NAIS} designs an attention network for interpretable collaborative filtering, 
which can identify significant items in the interaction sequence for prediction. 
RUM~\cite{DBLP:conf/wsdm/ChenXZT0QZ18/RUM} develops user memory networks for sequential recommendation, 
extracting patterns to show how previous items affect future user actions. 
Although these models offer good transparency, 
they are limited to capturing low-order relationships and fail to incorporate sequence information, resulting in subpar performance in sequence recommendation tasks (see Table~\ref{tbl:performance-cmp}).
Recently, self-attention-based models~\cite{DBLP:conf/icdm/KangM18/SASRec, DBLP:conf/cikm/SunLWPLOJ19/Bert4Rec, DBLP:conf/www/FanLWWNZPY22/STOSA, DBLP:conf/ijcai/ChenZJYLP0H023/PMAN} gain prominence and demonstrate strong performance on sequential data. 
By visualizing attention maps, these models reveal item dependencies, highlighting their interpretability. 
However, several studies~\cite{DBLP:conf/acl/SerranoS19/AttentionNotExp, DBLP:conf/naacl/JainW19/AttentionNotExp2} note that artificially removing key weights often does not impact the output distributions, 
raising concerns that attention weight distribution may not accurately reflect the model's actual decision-making process.
In contrast, our approach transparently captures the contributions of both low-order and high-order patterns, 
enhancing interpretability while achieving outstanding model performance.

Aspect-based approaches, in contrast, often integrate external information 
(e.g., user reviews, item attributes, item relationships) to achieve fine-grained interpretability. 
EFM~\cite{DBLP:conf/sigir/ZhangL0ZLM14/EFM} analyzes user reviews to extract aspects of item and user perspectives, 
generating explanations through a matrix factorization-based model. 
TriRank~\cite{DBLP:conf/cikm/HeCKC15/TriRank} similarly analyzes item aspects from user reviews 
and models the user-item-aspect relationship using tripartite graphs, achieving explainable collaborative filtering. 
PGPR~\cite{DBLP:conf/sigir/XianFMMZ19/PGPR} constructs graphs using item attributes and relational features, 
employing reinforcement learning to reason about target items, 
with the reasoning paths serving as explanations. 
NS-ICF~\cite{DBLP:conf/www/ZhangYWW22/NS-ICF} designs fully transparent three-tower structures based on neuro-rule networks~\cite{WangZLW24,Zhang-IJCAI23}
for attribute-based recommendations based on neuro-symbolic methods, 
generating highly interpretable rules and corresponding weights. 
CountER~\cite{DBLP:conf/cikm/TanXG00Z21/CountER} relies on counterfactual reasoning, 
where counterfactual explanations reveal which aspect changes affect item recommendations. 
These methods' explainability relies on rich and accurate aspect information; 
however, limited data may impact their effectiveness.

This work attempts to address the limitation of similarity-based methods in modeling higher-order item relationships 
by introducing probabilistic operators and only item ID is considered as input.
It is beneficial for providing multi-level interpretability while achieving high accuracy.

\begin{figure*}
    \centering
    \includegraphics[width=1.0\linewidth]{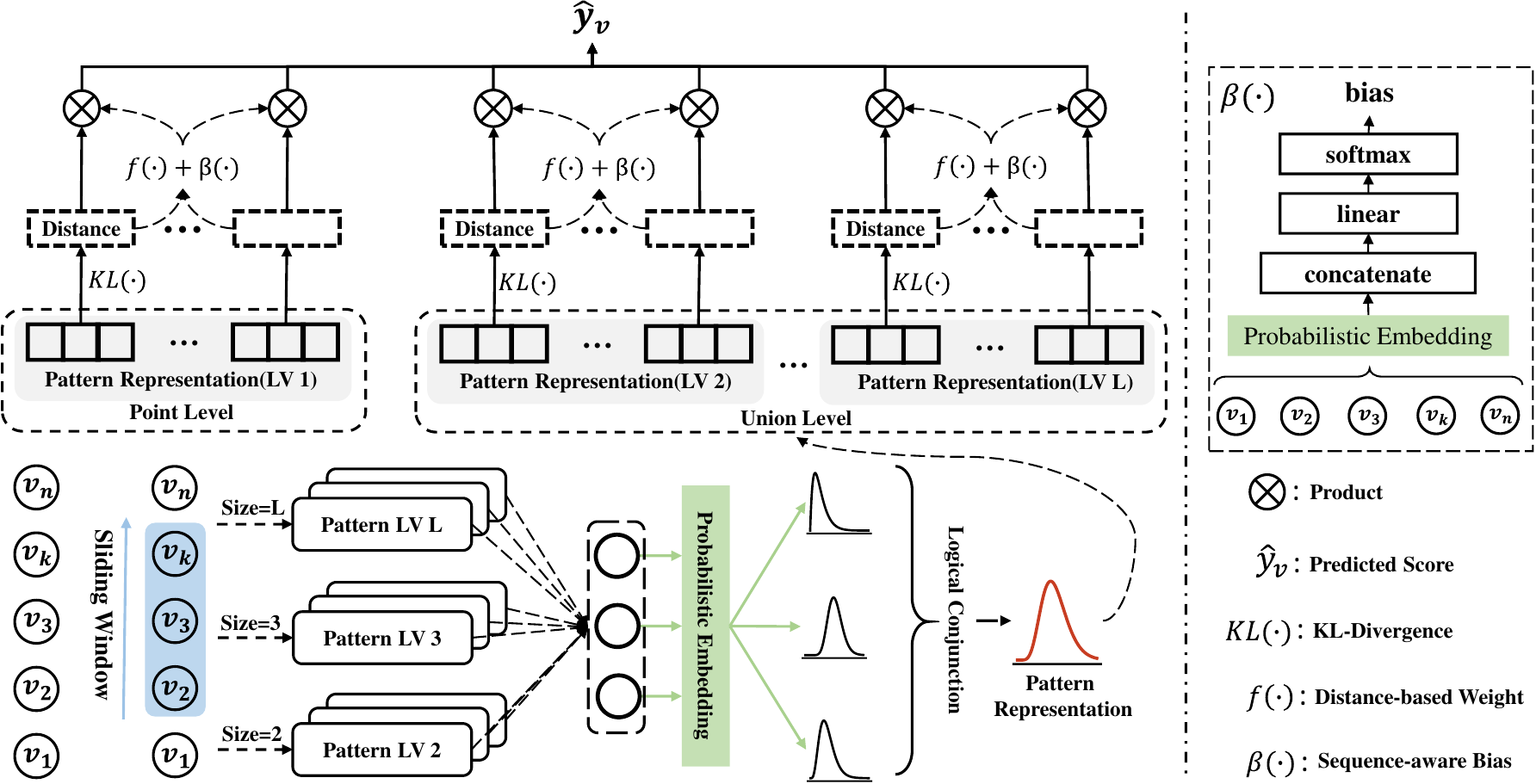}
    \caption{
    Architecture of PTSR: The framework employs multi-scale sliding windows for pattern extraction, 
    utilizes probabilistic embeddings and logical operators for pattern representation, 
    and computes target scores through KL-Divergence with integrated weights and biases.
    }
    \label{fig:PTSR}
\end{figure*}

\subsection{Novel Embedding for Recommendation}
Recently, novel types of embedding, such as distribution-based and geometry-based, etc., 
have begun to demonstrate remarkable performance. 
These embeddings differ fundamentally from traditional embedding 
by not confining users or items to a single point in vector space. 
Several studies have sought to integrate these advancements into recommendation systems to enhance model performance and cognitive capabilities.
For instance, DDN~\cite{DBLP:conf/sigir/ZhengLLZY19/DDN}, 
PMLAM~\cite{DBLP:conf/kdd/MaMZTLC20/PMLAM}, 
DT4SR~\cite{DBLP:conf/cikm/FanL00Y21/DT4SR}, 
and STOSA~\cite{DBLP:conf/www/FanLWWNZPY22/STOSA} have challenged the efficacy of the dot product, 
which is argued to violate the triangular inequality and lead to sub-optimal results. 
To address this, they propose modeling users and items as Gaussian distributions. 
Apart from those representations, other studies have explored Beta~\cite{DBLP:conf/nips/RenL20/BetaE} and 
Gamma~\cite{DBLP:conf/emnlp/YangQLLL22/GammaE} distribution-based representations, 
introducing logical operators customized for closures that operate upon these distributions, 
thereby laying the groundwork for subsequent methodologies. 
For example, SRPLR~\cite{DBLP:conf/ijcai/YuanZX0LS023/SRPLR} employs Beta embedding as foundational representations 
and merges its logical operators with neural networks to amplify the efficacy of traditional models.
Beyond distribution types, geometry-based embedding gains popularity as well. 
Zhang et al.~\cite{DBLP:conf/wsdm/ZhangLZHZLZHO21} propose the use of hypercuboids 
to explicitly represent and model user interests. 
CBox4CR~\cite{DBLP:conf/sigir/LiangZDXLY23/CBox4CR} seeks to refine model cognition through Box embedding.
This enables logical operations on their closures.

To harness the strong representational powers of these novel embeddings and the logical operators established upon them, 
we integrate probabilistic embeddings (e.g., Beta and Gamma embeddings) into our model, 
which facilitate capturing item relationships and modeling the pattern representations introduced later.

\section{Methodology}\label{sec:metho}

In this section, we first briefly introduce the task of interpretable sequential recommendation,
followed by details of Pattern-wise Transparent Sequential Recommendation (PTSR).
In addition, the optimization objective of PTSR is succinctly clarified 
and the complexity of the model will be analyzed.

\subsection{Problem Formulation}

Given a user's historical interaction sequence $s = [v_1, v_2, \ldots, v_n]$, 
where $n$ is the maximum sequence length,
sequential recommendation aims to predict the next item $v_{n+1}$ that the user is most likely to click or purchase.
Specifically, PTSR will score each candidate item to measure how well it matches the user's historical interactions.
Then these scores can be used to filter out the most relevant items from the candidates.

Apart from recommendation precision,
there are additional requirements for an interpretable sequence model; 
that is, key items or item combinations belonging to the user interaction sequence $s$ 
should be identified to account for the model's predictions.
This poses a challenge to traditional approaches due to their black-box nature.
In this paper, PTSR aims to identify a segment of the sequence (dubbed pattern) as the justification.
The items within this pattern collectively explain why a high score is assigned to the recommended item.

\subsection{Model Details of PTSR}

Fig.~\ref{fig:PTSR} illustrates the model architecture of PTSR including three modules:
\begin{itemize}[leftmargin=*]
    \item[1.]
    \textbf{Multi-level Pattern Extraction.}
    Sliding windows of different sizes are first employed to extract fine-grained patterns 
    by traversing the interaction sequence from left to right.
    Every pattern contains one or several consecutive item IDs.
    \item[2.]
    \textbf{Pattern Representation Modeling.}
    For each pattern, PTSR transforms its item IDs into probabilistic embedding,
    and then conjoins the items using probabilistic conjunction~\cite{DBLP:conf/nips/RenL20/BetaE, DBLP:conf/emnlp/YangQLLL22/GammaE}
    to obtain the representation of the pattern .
    \item[3.]
    \textbf{Pattern Contribution Estimation.}
    Firstly, the distance between each representation of the pattern and the representation of the target item can be calculated by KL-Divergence.
    A weight unit then uses a softmax function to transform negative distance into non-normalized
    weights in conjunction with a sequence-aware bias.
    They are combined with distances to determine each pattern's contribution.
\end{itemize}

\subsubsection{Multi-level Pattern Extraction}

Previous methods primarily focus on the significance of low-order (point-level) item relationships to maintain decision transparency~\cite{DBLP:conf/wsdm/ChenXZT0QZ18/RUM, DBLP:journals/tkde/HeHSLJC18/NAIS}. 
Additionally, other approaches that account for higher-order (union-level) item relationships typically achieve this by stacking multiple neural network modules~\cite{DBLP:conf/icdm/KangM18/SASRec, DBLP:conf/cikm/SunLWPLOJ19/Bert4Rec, DBLP:conf/www/FanLWWNZPY22/STOSA}. 
However, these higher-order relationships often lack interpretability due to the nonlinear functions involved in these modules.
To model point-level and union-level item relationships while preserving model transparency,
we begin by extracting multi-level patterns from the sequence directly, 
which form the \textbf{atomic units} for the subsequent transparent decision-making process.
To be more specific, 
it explicitly splits the sequence $s=[v_1, v_2, \ldots, v_n]$ via a sliding window of size $l$,
and the $k$-th segment is defined as follows:
\begin{equation*}
    p_k^{(l)} = [v_k, v_{k+1}, \ldots, v_{k+l-1}], \quad k=1,2, \ldots, n - l + 1.
\end{equation*}
Notably, patterns of different levels should be understood from different perspectives \cite{DBLP:conf/wsdm/TangW18/Caser}:
\begin{itemize}[leftmargin=*]
  \item $p^{(1)}$ inherently is an individual item.
  Such \textit{point-level} pattern appears simple but is capable of providing the most straightforward and human-friendly explanation.
  For example, 
  the purchase of a camera rather than a T-shirt contributes more to the purchase of a memory card.
  A reliable sequence model should be able to accurately learn such similarities across items.

  \item
  Apart from the point-level pattern, 
  a sliding window of size $l > 1$ yields the \textit{union-level} pattern,
  which contains a group of items interacted by the user sequentially.
  These union-level patterns are of great value in capturing the user intent at a deeper level.
  For example, the co-occurrence of `phone' and `headphone' implies 
  a clear interest in electronics, also making it reasonable to recommend `memory card'.
  This higher-order interpretability enhances clarity in 
  understanding the combinational relationships between items. 
  By mining the combinational display of multiple items based on users' interactions, 
  it becomes easier to gain users' trust 
  compared to relying on isolated single-item similarities.
\end{itemize}
Since both point-level and union-level patterns are essential for a transparent decision-making process, we collect all patterns with levels ranging from 1 to $L$.
The remaining challenge lies in effectively integrating the items for pattern-wise representations.

\begin{figure}[!t]
    \centering
    \setlength{\abovecaptionskip}{0.cm}
    \includegraphics[width=\linewidth]{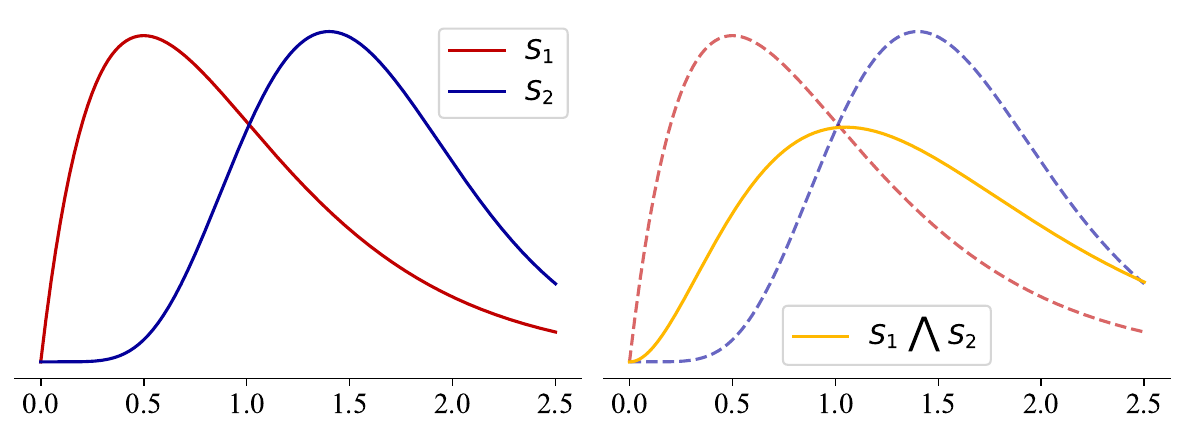}
    \caption{
        Illustration of probabilistic conjunction operator. 
        $S_1 \bigwedge S_2$ denotes the conjunction of distribution $S_1$ and $S_2$, 
        achieved by the weighted product of the probability density function (PDF).
    }
    \label{fig:gamma_and}
\end{figure}

\subsubsection{Pattern Representation Modeling}

It is straightforward to capture `common' characteristics for point-level patterns
but not so for union-level patterns.
Previous methods of relative transparency typically 
embed each item into a learnable vector $\mathbf{v} \in \mathbb{R}^d$,
and then the pattern representation is obtained in a weighted average manner:
\begin{align*}
  \mathbf{p}_k = \sum_{i=k}^{k+l-1} w_i \cdot \mathbf{v}_i \in \mathbb{R}^d,
\end{align*}
where $w_i$ is the weight associated with $\mathbf{v}_i$ 
(e.g., $w_i \equiv 1 / l$ for mean pooling).
However, this simple fusion of embeddings is still limited to similarity and cannot model complex relationships between items.

To address this issue, we resort to probabilistic embedding in the form of probability distributions (e.g., Gamma, Beta, and Gaussian).
Specifically, the elements in $\mathbf{v}$ are the parameters of probabilistic distribution.
For example, the Gamma distribution is jointly described 
by a shape parameter $\alpha>0$ and a scale parameter $\beta>0$,
while the Beta distribution is defined by two shape parameters $(\alpha, \beta)$.
Both of them yield the probabilistic embedding that resembles:
\begin{align}
  \mathbf{v} = (\bm{\alpha}, \bm{\beta}) = [(\alpha_1, \beta_1), (\alpha_2, \beta_2), \ldots, (\alpha_d, \beta_d)] \in \mathbb{R}_+^{2d}.
\end{align}
In this case, multiple independent distributions defined by different parameters 
can represent different characteristics of the item, 
which can enhance expressiveness
~\cite{DBLP:conf/emnlp/YangQLLL22/GammaE,DBLP:conf/nips/RenL20/BetaE}.
It has two distinct advantages over the typical vector representation:
1) The difference along the dimensions can be separately measured by probabilistic metrics 
such as Kullback-Leibler (KL) divergence instead of Euclidean distance.
KL-Divergence can naturally model the asymmetry and uncertainty that prevail among items.
2) Complex relationships between items can be modeled 
using logical operators (introduced next) based on probabilistic embeddings.

For aggregating item representations within patterns, 
unlike traditional operations such as pooling or GRU, 
we introduce a logical conjunction operator to more accurately model the relationships between items. 
PTSR primarily builds on BetaE~\cite{DBLP:conf/nips/RenL20/BetaE} and GammaE~\cite{DBLP:conf/emnlp/YangQLLL22/GammaE}\footnote{Other probability distributions supporting logical operators, such as Gaussian~\cite{DBLP:conf/nips/ChoudharyRKSR21/PERM} and Dirichlet~\cite{DBLP:conf/www/WangZSYY22/DiriE} distributions, are also applicable.}, 
where the conjunction among multiple distributions is implemented as the product of their probability density function (PDF).
PDF is denoted by $f(\cdot; \alpha, \beta)$,
the product of a collection of PDFs can be derived as follows:
\begin{align}
  \label{eq-scalar-fusion}
  \prod_{i} f^{w_i}(\alpha_i, \beta_i) 
  \propto f(\sum_{i} w_i \cdot \alpha_i, \sum_{i} w_i \cdot \beta_i).
\end{align}
This `weighted average' operation is a scalar form of the probabilistic conjunction operation \cite{DBLP:conf/nips/RenL20/BetaE},
which yields a PDF that measures how well the different distributions agree with each other (see Fig.~\ref{fig:gamma_and}).
Probability distributions offer greater expressiveness than real values, 
allowing pattern representations obtained through probabilistic conjunction 
to capture complex interaction relationships between items 
more effectively than standard weighted fusion. 
We substantiate this claim with comparative experiments presented in Section~\ref{sec:ablation}.

We are now ready to extract the conjunction relationships within a single pattern.
Given an $l$-length pattern $p_k=[v_k, v_{k+1}, \ldots, v_{k+l-1}]$,
a vectorized probabilistic conjunction then separately fuses items along the embedding dimension,
\begin{equation}
  \label{eq-vector-fusion}
  \mathbf{p}_k := \bigg(
    \sum_{i=k}^{k+l-1} \mathbf{w}_i \odot \bm{\alpha}^{v_i}, \sum_{i=k}^{k+l-1} \mathbf{w}_i \odot \bm{\beta}^{v_i}
    \bigg) \in \mathbb{R}_+^{2d},
\end{equation}
where $\odot$ denotes the element-wise product, 
and a self-attentive mechanism is used to boost the model's expressive power,
\begin{equation*}
  \mathbf{w}_i = \frac{
    \exp \big(
      \MLP (\bm{\alpha}^{v_i} \oplus \bm{\beta}^{v_i})
    \big)
  }{
    \sum_{j=k}^{k+l-1}
    \exp \big(
      \MLP (\bm{\alpha}^{v_j} \oplus \bm{\beta}^{v_j})
    \big)
  } \in \mathbb{R}^d,
\end{equation*}
where $\oplus$ denotes the vector concatenation.

\subsubsection{Pattern Contribution Estimation}
For a given candidate item $v$, 
the most straightforward way to measure the relevance between pattern $p_k^{(l)}$ and $v$ 
is to calculate the distance between their probabilistic embeddings using KL-Divergence.
And the smaller the distance, the higher the correlation between $p_k^{(l)}$ and $v$.
Specifically, it can be expressed as:
\begin{equation}
    Dis_v^{p_k^{(l)}} = 
            \sum_{i=1}^d \text{KL}
            \big(
                f(\alpha_i^{v}, \beta_i^{v}) \| f(\alpha_i^{p_k^{(l)}}, \beta_i^{p_k^{(l)}})
            \big)
\end{equation}

Therefore, the primary principle of pattern contribution learning 
is to keep the key patterns close to the target item.
However,
supervised pattern contribution learning is infeasible in practice due to the unavailability of ground-truth key patterns.
One might expect that 
minimizing the sum of distances from the target item
and simultaneously maximizing the distances from the negative item
would produce satisfactory results, 
but this is not the case for some reasons below.
\begin{itemize}[leftmargin=*]
  \item
  The distance (i.e., the KL-Divergence) between key patterns and the target item may decrease rapidly at the beginning of training.
  Soon it becomes easier to reduce the distance of non-critical patterns,
  eventually leading to marginal contribution differences across various patterns.
  \item
  Summing the distances is too na\"ive to achieve competitive recommendation performance.
  Two different candidate items may be considered equally good in this vein 
  despite the large difference in distance from the key pattern.
  For instance, consider a pattern $p_1$ whose distance to the candidate $v_1$ is 0.1 and to the candidate $v_2$ is 0.8. 
  In contrast, another pattern $p_2$ has a distance of 4.9 to $v_1$ and 4.2 to $v_2$.
  Compared to $p_2$, $p_1$ is arguably the key pattern for both candidate items,
  and thus $v_1$ should be recommended as it receives more support from $p_1$.
  But they are of the same value according to the summing distance.
\end{itemize}
Hence, it is necessary to correct the total distance for next-item recommendation 
to highlight the key patterns that have low distances from the target item.

\textbf{Distance-based weight.}
Patterns deemed important will exhibit smaller distances from the target item 
and thus require larger weights to emphasize their contribution. 
Conversely, patterns far from the target item are of lower importance and should be assigned with smaller weights.
For maintaining a proportional correspondence between the distance values and the weight values, 
the distance-based weights are normalized through a softmax function over the negative distances.
For example, the weight of the $k$-th pattern at level $l$ can be expressed as follows:
\begin{equation}
    \eta_v^{p_k^{(l)}} = 
    \frac{\exp(-Dis_v^{p_k^{(l)}})}{\sum_{j=1} \exp (-Dis_v^{p_j^{(l)}})}.
\end{equation}

It is worth noting that the distance-based weights are independently calculated for each level.
In this vein, 
patterns at the same level will compete with each other, 
and eventually, the most important patterns will stand out.
Besides, patterns at different levels will collaborate to improve recommendation performance.

\textbf{Sequence-aware bias.}
The distance-based weight can not only discern the importance of patterns 
but also detect alterations in the sequence of items, 
a crucial attribute in sequential recommendation. 
However, there is one exception to the incapacity to perceive changes in the sequence's order --- when the entire sequence is reversed. 
The primary reason for this exception is that, in cases where only some items change their order, there must exist a pair of items that were not adjacent before but are now adjacent. 
In such instances, a sliding window of size 2 can capture this change. 
Nevertheless, when the entire sequence is reversed, there is no alteration in the adjacency between items.

We opt to introduce a sequence-aware bias to the distance-based weight, allowing it to adapt as the overall sequence order changes. 
Its practical significance lies in its interpretability as an evolutionary direction of user interest. 
Specifically, we employ an $\MLP$ to derive the sequence-aware weight, leveraging the sensitivity of the $\MLP$ to input order~\cite{DBLP:conf/ijcai/LiZL0WG22/MLP4Rec}. 
Concerning the input for the $\MLP$, for each item in the sequence, we initially combine $\alpha$ and $\beta$ in their Gamma embedding using $\frac{\alpha}{\alpha + \beta}$. 
The updated representation of the $i$-th item can be expressed as $\mathbf{e}_i=\frac{\bm{\alpha}_i}{\bm{\alpha}_i + \bm{\beta}_i}$. 
This operation directly reduces the number of parameters in the MLP by half while preserving information compared to directly concatenating $\alpha$ and $\beta$. 
Subsequently, we concatenate these updated representations of all items to form the input for the $\MLP$.
For $\MLP$'s output, we also use softmax for normalization.
The sequence-aware bias of $k$-th pattern at level $l$ is expressed as follows:
\begin{equation}
    \delta_k^{(l)} = \big[\mathrm{Softmax}(\MLP^{(l)}(\mathbf{e}_1\oplus \cdots \oplus \mathbf{e}_n)) \big]_k\,,
\end{equation}
where $n$ is the sequence length.
Given that each level contains a distinct number of patterns, we assign a unique bias to each level accordingly.

Combining the distance-based weight and sequence-aware bias, we can get the final pattern correction weight $\eta_v^{p_k^{(l)}} + \lambda \delta_k^{(l)}$
where $\lambda$ is a hyperparameter used to control the effect of bias.

\subsection{Prediction and Training Objective}
With the pattern contribution,
the eventually corrected score for prediction is defined as follows:
\begin{equation}
    \hat{y}_v = \sum\limits_{l=1}^{L} \cdot \Bigg [\sum\limits_{k=1}^{n-l+1} \big(\eta_v^{p_k^{(l)}} + \lambda \delta_k^{(l)} \big) \cdot (\gamma - Dis_v^{p_k^{(l)}}) \Bigg].
    \label{eq:score}
\end{equation}
where \( \eta_v^{p_k^{(l)}} \) and \( \delta_k^{(l)} \) denote the distance-based weight and sequence-aware bias of the \( k \)-th pattern in the \( l \)-th level, respectively. \( \lambda \) adjusts the relative influence, and \( \gamma \) controls the distance optimization range.

Note that $\hat{y}_v$ differentially integrates contributions across patterns and levels.
As a result, a high prediction score requires a strong agreement among the majority of key patterns,
and thus key patterns can be implicitly learned even in the absence of ground-truth labels.
In addition, $\gamma$ is a margin hyperparameter to prevent $\hat{y}_v$ from always being less than or equal to 0 during training.

The training objective of PTSR is the binary cross-entropy loss, 
which is designed to minimize the corrected score $\hat{y}_{v_+}$ to target item $v_+$,
while simultaneously maximizing the corrected score $\hat{y}_{v_-}$ to negative items $v_-$ (uniformly sampling one negative sample for each positive item);
that is,
\begin{equation}
\begin{aligned}
    \mathcal{\ell}= 
    -\log \sigma (\hat{y}_{v_+})
    -\log\sigma (-\hat{y}_{v_-}).
\end{aligned}
\end{equation}
where $\sigma(\cdot)$ is the sigmoid function.

\subsection{Discussion}
In this part, we discuss the relationship between different levels of interpretability and analyze the computational complexity of PTSR.

\textbf{Interpretability.}
In PTSR, point-level and union-level interpretability assess item roles from different perspectives. 
Point-level focuses on individual items; 
a high correlation with the target item leads to salient score contributions. 
In contrast, union level considers an item and its surrounding neighbors; 
if relationships are lacking, it indicates an ineffective pattern, resulting in low impact.
Patterns at different levels can assess item relevance from various perspectives. 
Consider a target item where an item in the sequence exhibits low relevance at the point level but high relevance at the union level. 
This discrepancy may arise because the relevant features constitute only a small fraction of the total features, 
making them difficult to capture individually. 
However, if both the item and its neighbors share these relevant features, 
the union level can leverage their commonality to amplify these features, leading to higher relevance.
Therefore, compared to point-level, 
union-level primarily enhances certain interests by leveraging synergistic relationships between items.
When PTSR needs to discern such differences and decide whether to recommend a target item, 
the absence of negative correlation, meaning no cancellation, 
allows it to focus solely on which pattern exhibits stronger relevance, 
indicated by smaller distances or greater weights. 
Additionally, the magnitude of the bias corresponding to the pattern's position is also a factor to consider.

\textbf{Computational complexity.}
For each pattern of length $l$, 
the complexity for the probabilistic conjunction operation is approximately $\mathcal{O}(ld^2)$.
Since there are a total of $n-l+1$ patterns per level, 
the main computational complexity of PTSR is $\mathcal{O}(nL^2d^2)$.
Recall that a single self-attention module requires $\mathcal{O}(nd^2 + n^2d)$ overhead.
Since $L=2,3$ is often the optimal value regardless of the performance  (see Section \ref{sec:ablation}) or interpretability point of view,
the efficiency of PTSR therefore is comparable to that of normal attention-based methods.

\begin{table}[]
\caption{Dataset statistics after preprocessing.}
\label{tab:data_sta}
\renewcommand\arraystretch{1.3}
\begin{tabular}{cccccc}
\toprule
Dataset & \#Users & \#Items & \#Interactions & Avg. Len. & Sparsity \\ \midrule
Beauty  & 22,363   & 12,101   & 198,502        & 8.87    & 99.93\%  \\
Toys    & 19,412  & 11,924  & 167,597       & 8.63    & 99.99\%   \\
Tools   & 16,638   & 10,217   & 134,476        & 8.08    & 99.92\%  \\
Yelp    & 30,431   & 20,033   & 316,354        & 10.39   & 99.95\%  \\   
Sports    & 331,844   & 103,911   & 2,835,125        & 8.54   & 99.99\%  \\   \bottomrule
\end{tabular}
\end{table}

\section{Experiments}\label{sec:exp}
This section introduces extensive experiments to address the following five research questions:
\begin{itemize}[leftmargin=2.5em]
    \item[\textit{RQ1}] How is the recommendation performance of PTSR compared to other sequence models?
    \item[\textit{RQ2}] To what extent do specific components (e.g., Weight, Bias) contribute to the effectiveness of PTSR?
    \item[\textit{RQ3}] What is the performance associated with varying levels of patterns?
    \item[\textit{RQ4}] How critical is the hyperparameter $\lambda$ to the performance of PTSR?
    \item[\textit{RQ5}] How interpretable is PTSR in comparison to other methods?
    \item[\textit{RQ6}] How is the training stability of PTSR?
\end{itemize}

\begin{table*}[!t]
\centering
\caption{
Overall performance comparison across five datasets. 
The best results among all methods are marked in \textbf{bold},
while the second best results among the baselines are \underline{underlined}.
H@K and N@K represent Hit Ratio (HR) and Normalized Discounted Cumulative Gain (NDCG), respectively.
`Improv.' represents the relative improvement over the best baseline.
Paired t-test is performed over 5 independent runs for evaluating $p$-value
(* indicates statistical significance with a $p$-value $< 0.01$).
PTSR-B/G indicates the use of Beta/Gamma embedding.
}
\renewcommand\arraystretch{1.3}
\label{tbl:performance-cmp}
\setlength{\tabcolsep}{0.8mm}{
\begin{tabular}{llccccccccccccccc}
\toprule
\multicolumn{1}{c}{\multirow{2}{*}{\rotatebox{90}{Dataset}}} & \multirow{2}{*}{Metric} & \multicolumn{3}{c}{Traditional} & \multicolumn{3}{c}{Transparent} & \multicolumn{2}{c}{Attention-Based} & \multicolumn{3}{c}{Novel Embedding}  & \multicolumn{4}{c}{Ours}            \\
\cmidrule(lr){3-5} \cmidrule(lr){6-8}\cmidrule(lr){9-10}\cmidrule(lr){11-13}\cmidrule(lr){14-17}
\multicolumn{1}{c}{}                         &                         & Caser    & GRU4Rec   & MSGIFSR  & NAIS      & RUM      & PMAN     & SASRec        & BERT4Rec            & STOSA  & CBox4CR      & SRPLR        & PTSR-B & Improv. & PTSR-G & Improv. \\ \midrule
\multirow{4}{*}{\rotatebox{90}{Beauty}}                      & H@5                     & 0.3206   & 0.3402    & 0.3598   & 0.3256    & 0.3763   & 0.3514   & 0.3882        & 0.3989              & 0.3725 & 0.3997       & {\ul 0.4100} & 0.4492* & 9.54\%  & {\textbf{0.4512}}* & 10.05\% \\
                                             & H@10                    & 0.4252   & 0.4410    & 0.4475   & 0.4240    & 0.4764   & 0.4508   & 0.4832        & 0.4987              & 0.4772 & {\ul 0.5044} & 0.5024       & 0.5532* & 9.67\%  & {\textbf{0.5573}}* & 10.49\% \\
                                             & N@5                     & 0.2303   & 0.2511    & 0.2783   & 0.2379    & 0.2779   & 0.2589   & 0.2963        & 0.3039              & 0.2737 & 0.2959       & {\ul 0.3219} & 0.3395* & 5.47\%  & {\textbf{0.3404}}* & 5.75\%  \\
                                             & N@10                    & 0.2641   & 0.2836    & 0.3066   & 0.2696    & 0.3102   & 0.2910   & 0.3270        & 0.3361              & 0.3075 & 0.3297       & {\ul 0.3517} & 0.3731* & 6.08\%  & {\textbf{0.3747}}* & 6.54\%  \\ \midrule
\multirow{4}{*}{\rotatebox{90}{Toys}}                        & H@5                     & 0.3014   & 0.3308    & 0.3323   & 0.2935    & 0.3420   & 0.3463   & 0.3867        & 0.3734              & 0.3816 & 0.3791       & {\ul 0.4076} & 0.4291* & 5.27\%  & {\textbf{0.4302}}* & 5.54\%  \\
                                             & H@10                    & 0.4009   & 0.4367    & 0.4244   & 0.3876    & 0.4409   & 0.4503   & 0.4925        & 0.4786              & 0.4795 & 0.4848       & {\ul 0.5035} & 0.5344* & 6.14\%  & {\textbf{0.5347}}* & 6.20\%  \\
                                             & N@5                     & 0.2179   & 0.2392    & 0.2493   & 0.2161    & 0.2533   & 0.2518   & 0.2866        & 0.2799              & 0.2898 & 0.2775       & {\ul 0.3186} & 0.3256* & 2.20\%  & {\textbf{0.3276}}* & 2.82\%  \\
                                             & N@10                    & 0.2500   & 0.2734    & 0.2790   & 0.2464    & 0.2853   & 0.2855   & 0.3208        & 0.3139              & 0.3213 & 0.3116       & {\ul 0.3495} & 0.3597* & 2.92\%  & {\textbf{0.3614}}* & 3.40\%  \\ \midrule
\multirow{4}{*}{\rotatebox{90}{Tools}}                       & H@5                     & 0.2188   & 0.2350    & 0.2221   & 0.2349    & 0.2618   & 0.2595   & 0.2891        & {\ul 0.3017}        & 0.2905 & 0.2758       & 0.2836       & {\textbf{0.3425}}* & 13.52\% & 0.3377* & 11.93\% \\
                                             & H@10                    & 0.3135   & 0.3414    & 0.2741   & 0.3266    & 0.3632   & 0.3596   & 0.3885        & {\ul 0.4067}        & 0.3897 & 0.3832       & 0.3813       & {\textbf{0.4534}}* & 11.48\% & 0.4485* & 10.28\% \\
                                             & N@5                     & 0.1520   & 0.1618    & 0.1971   & 0.1679    & 0.1863   & 0.1869   & 0.2107        & {\ul 0.2185}        & 0.2120 & 0.1943       & 0.2094       & {\textbf{0.2495}}* & 14.19\% & 0.2458* & 12.49\% \\
                                             & N@10                    & 0.1824   & 0.1961    & 0.1874   & 0.1974    & 0.2190   & 0.2184   & 0.2427        & {\ul 0.2523}        & 0.2440 & 0.2289       & 0.2408       & {\textbf{0.2853}}* & 13.08\% & 0.2815* & 11.57\% \\ \midrule
\multirow{4}{*}{\rotatebox{90}{Yelp}}                        & H@5                     & 0.5357   & 0.5661    & 0.6101   & 0.5275    & 0.6075   & 0.5174   & 0.5986        & 0.6167              & 0.5695 & {\ul 0.6383} & 0.6247       & {\textbf{0.6677}}* & 4.61\%  & 0.6651* & 4.20\%  \\
                                             & H@10                    & 0.6988   & 0.7322    & 0.7638   & 0.7023    & 0.7660   & 0.6727   & 0.7639        & 0.7599              & 0.7329 & {\ul 0.8026} & 0.7797       & {\textbf{0.8138}}* & 1.40\%  & 0.8107* & 1.01\%  \\
                                             & N@5                     & 0.3772   & 0.3998    & 0.4499   & 0.3711    & 0.4459   & 0.3711   & 0.4339        & 0.4567              & 0.4117 & {\ul 0.4622} & 0.4609       & 0.5013* & 8.46\%  & {\textbf{0.5025}}* & 8.72\%  \\
                                             & N@10                    & 0.4301   & 0.4537    & 0.4998   & 0.4277    & 0.4973   & 0.4214   & 0.4875        & 0.5032              & 0.4646 & {\ul 0.5156} & 0.5113       & 0.5487* & 6.42\%  & {\textbf{0.5498}}* & 6.63\%  \\ \midrule
{}                         & {H@5}                   & {0.3911} & { 0.2910} & { 0.4769} & { 0.3949} & { 0.4220} & { 0.4458} & { 0.4968} & { \ul 0.4995} & { 0.4665} & { 0.4890} & { 0.4940} & { 0.5125*} & { 3.16\%} & { \textbf{0.5130*}} & { 3.26\%} \\
{ }                         & { H@10}                  & { 0.5014} & { 0.3956} & { 0.5830} & { 0.5050} & { 0.5438} & { 0.5584} & { \ul 0.6196} & { 0.6115} & { 0.5840} & { 0.6115} & { 0.6166} & { 0.6317*} & { 1.95\%} & { \textbf{0.6321*}} & { 2.02\%} \\
{ }                         & { N@5}                   & { 0.2928} & { 0.2072} & { 0.3734} & { 0.2936} & { 0.3145} & { 0.3475} & { 0.3811} & { 
\ul 0.3943} & { 0.3605} & { 0.3750} & { 0.3805} & { \textbf{0.3990*}} & { 1.19\%} & { 0.3978*} & { 0.89\%} \\
\multirow{-4}{*}{\rotatebox{90}{ Sports}} & { N@10}                  & { 0.3284} & { 0.2410} & { 0.4077} & { 0.3291} & { 0.3539} & { 0.3838} & { 0.4208} & { \ul 0.4305} & { 0.3985} & { 0.4128} & { 0.4201} & { \textbf{0.4388*}} & { 1.93\%} & { 0.4363*} & { 1.35\%} \\ \bottomrule
\end{tabular}
}
\end{table*}

\subsection{Experimental Setup}

\textbf{Datasets.}
We evaluate our model on five publicly available datasets from two different sources:
\begin{itemize}[leftmargin=*]
    \item \textbf{Amazon} records user reviews of the site's products.
    The data is divided into multiple datasets according to the category of items.
    We select \textbf{Beauty}, \textbf{Toys}, \textbf{Tools}, and \textbf{Sports} that are known for high data sparsity.
    \item \textbf{Yelp} is a famous merchant review website, 
    by which the dataset released contains data on user ratings of merchants on the site. 
\end{itemize}

Following \cite{DBLP:conf/icdm/KangM18/SASRec, DBLP:conf/recsys/HeKM17/TransRec}, 
we filter out the users and items with fewer than 5 interactions,
and split the dataset in a \textit{leave-one-out} fashion.
The second last item is used for validation and the last item serves for testing.
The statistics of the processed datasets have been summarized in Table \ref{tab:data_sta}.

\textbf{Evaluation metrics.}
We employ two commonly used metrics for model evaluation, 
including Normalized Discounted Cumulative Gain (NDCG) and Hit Ratio (HR).
NDCG is a metric assessing the efficiency of a ranking system by considering the placement of relevant items within the ranked list. 
HR is the accuracy of ground-truth items that appear in top $N$ recommendation list.
To enhance the evaluation's effectiveness, we employ \textit{real-plus-N}~\cite{DBLP:conf/recsys/SaidB14/Real-plus-N} to calculate the metrics. 
More specifically, we randomly choose 100 items that users have not interacted with as negative samples. These are then combined with the ground truth to create a candidate set for ranking.
NDCG@5, NDCG@10, HR@5, and HR@10 are reported.

\subsection{Baselines}
In order to fully demonstrate the effect of our model, 
we choose four different groups of recommendation baselines.
The first group employs traditional methods, such as recurrent neural networks, convolutional networks, and graph networks, to model user sequences:

\begin{itemize}[leftmargin=*]
    \item \textbf{GRU4Rec~\cite{DBLP:journals/corr/HidasiKBT15/GRU4Rec}:} Early proposal of using GRU to model user action sequences, mainly applied to session-based recommendation.
    \item \textbf{Caser~\cite{DBLP:conf/wsdm/TangW18/Caser}:} A method that treats user behavior sequences as `image' and use convolution kernels to learn sequential patterns.
    \item \textbf{MSGIFSR~\cite{DBLP:conf/wsdm/Guo0SZWBZ22/MSGIFSR}:} The method proposes using graph attention network to model the relationships between user interests at different granularity.
\end{itemize}

The second group of models features a transparent structure, providing good interpretability:
\begin{itemize}[leftmargin=*]
    \item \textbf{NAIS~\cite{DBLP:journals/tkde/HeHSLJC18/NAIS}:} Through a more transparent attention network, it is capable of distinguishing which historical items in a user profile 
    are more important for a prediction.
    \item \textbf{RUM~\cite{DBLP:conf/wsdm/ChenXZT0QZ18/RUM}:} It introduces a memory mechanism to dynamically manage user interaction records 
    and capture the importance of items in the interaction history.
    \item \textbf{PMAN~\cite{DBLP:conf/ijcai/ChenZJYLP0H023/PMAN}:} This method introduces a mask to reduce noise during self-attention modeling while removing the feed-forward network, increasing the transparency of model.
\end{itemize}
The methods in the third group leverage the self-attention mechanism to model long-term behavioral patterns:
\begin{itemize}[leftmargin=*]
    \item \textbf{SASRec~\cite{DBLP:conf/icdm/KangM18/SASRec}:} A classic model in sequential recommendation, applying a left-to-right attention mask to capture user's behavior.
    \item \textbf{BERT4Rec~\cite{DBLP:conf/cikm/SunLWPLOJ19/Bert4Rec}:} A method using a bidirectional self-attention mechanism, introducing a cloze task to predict masked items.
\end{itemize}

\begin{table*}[!t]
\setlength{\tabcolsep}{1.2mm}
\renewcommand\arraystretch{1.3}
\caption{
Ablation study(NDCG@5, NDCG@10, HR@5, HR@10) on four datasets,
with best effects highlighted in bold.
}
\label{tbl:ablation}
\begin{tabular}{lcccccccccccccccc}
\toprule
\multirow{2}{*}{Architecture} & \multicolumn{4}{c}{Beauty}                                            & \multicolumn{4}{c}{Toys}                                              & \multicolumn{4}{c}{Tools}                                             & \multicolumn{4}{c}{Yelp}                                              \\ \cmidrule(lr){2-5} \cmidrule(lr){6-9} \cmidrule(lr){10-13} \cmidrule(lr){14-17}
                           & N@5          & N@10         & HR@5            & HR@10           & N@5          & N@10         & HR@5            & HR@10           & N@5          & N@10         & HR@5            & HR@10           & N@5          & N@10         & HR@5            & HR@10           \\ \midrule
Default                    & \textbf{0.3404} & \textbf{0.3747} & \textbf{0.4512} & \textbf{0.5573} & \textbf{0.3276} & \textbf{0.3614} & \textbf{0.4302} & \textbf{0.5347} & \textbf{0.2458} & \textbf{0.2815} & \textbf{0.3377} & \textbf{0.4485} & \textbf{0.5025} & \textbf{0.5498} & \textbf{0.6651} & \textbf{0.8107} \\ 
w/o W                   & 0.3007          & 0.3337          & 0.4008          & 0.5030          & 0.2968          & 0.3306          & 0.3918          & 0.4963          & 0.1718          & 0.2089          & 0.2537          & 0.3687          & 0.4155          & 0.4725          & 0.5888          & 0.7644          \\
w/o B                   & 0.3228          & 0.3593          & 0.4376          & 0.5502          & 0.2980          & 0.3355          & 0.4040          & 0.5199          & 0.2398          & 0.2754          & 0.3357          & 0.4456          & 0.4979          & 0.5454          & 0.6632          & 0.8094          \\
w/o W + B               & 0.2490          & 0.2832          & 0.3428          & 0.4489          & 0.2204          & 0.2548          & 0.3093          & 0.4163          & 0.1672          & 0.2003          & 0.2399          & 0.3428          & 0.4131          & 0.4646          & 0.5664          & 0.7257          \\
Replace KL                 & 0.2397          & 0.2720          & 0.3317          & 0.4317          & 0.1879          & 0.2208          & 0.2620          & 0.3640          & 0.1845          & 0.2165          & 0.2592          & 0.3588          & 0.3423          & 0.4029          & 0.4911          & 0.6784          \\
Replace ProbE              & 0.2163          & 0.2533          & 0.3106          & 0.4247          & 0.1908          & 0.2294          & 0.2697          & 0.3896          & 0.1737          & 0.2077          & 0.2502          & 0.3556          & 0.3398          & 0.3973          & 0.4808          & 0.6592          \\ \bottomrule
\end{tabular}
\end{table*}

\begin{figure*}[!t]
    \centering
    \subfigure[Beauty]{
        \begin{minipage}[t]{0.245\linewidth}
        \centering
        \includegraphics[width=1.8in]{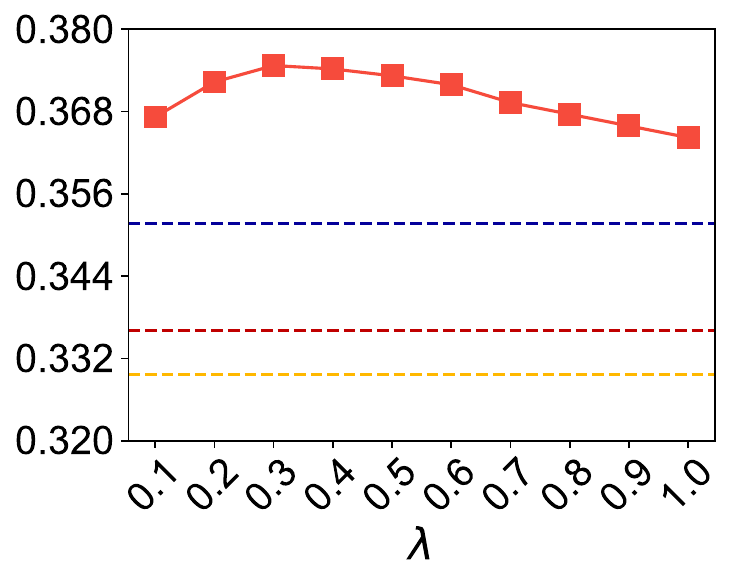}
        \end{minipage}%
    }%
    \subfigure[Toys]{
        \begin{minipage}[t]{0.245\linewidth}
        \centering
        \includegraphics[width=1.8in]{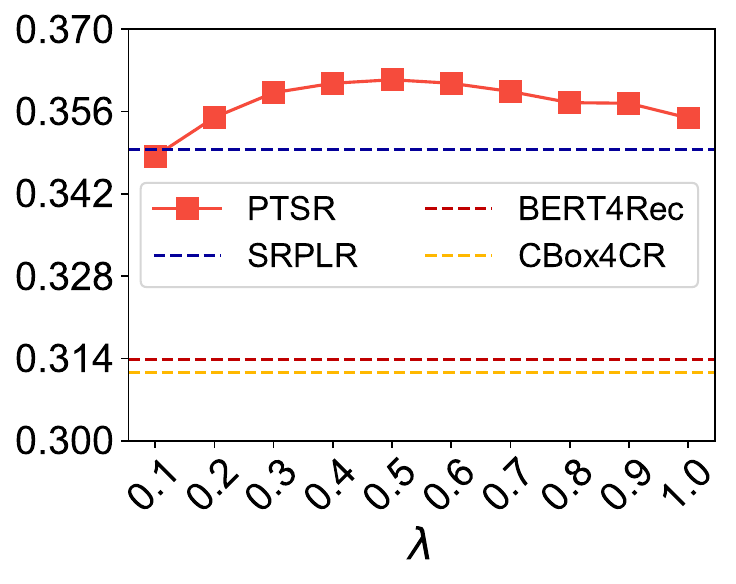}
        \end{minipage}%
    }%
    \subfigure[Tools]{
        \begin{minipage}[t]{0.245\linewidth}
        \centering
        \includegraphics[width=1.8in]{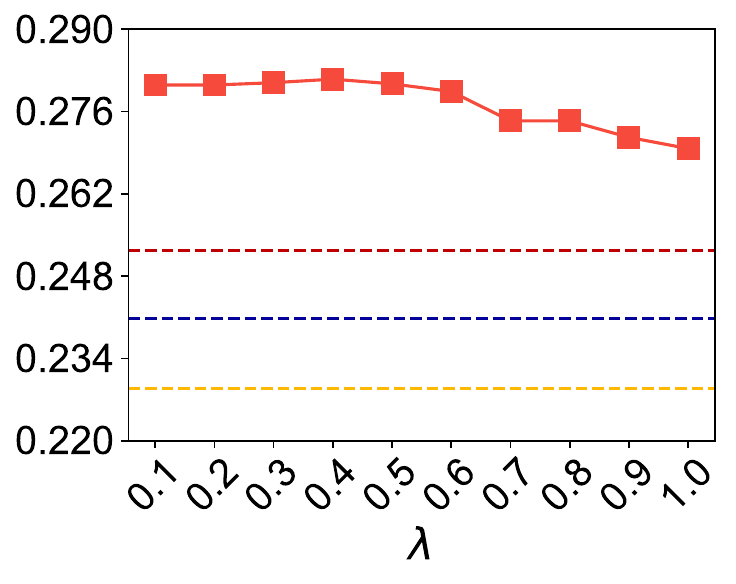}
        \end{minipage}
    }%
    \subfigure[Yelp]{
        \begin{minipage}[t]{0.245\linewidth}
        \centering
        \includegraphics[width=1.8in]{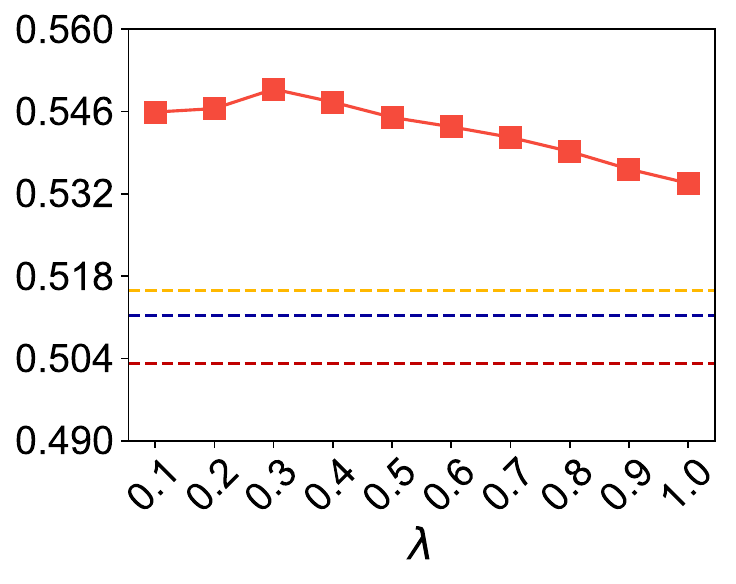}
        \end{minipage}
    }%
    \centering
    \caption{NDCG@10 performance over various $\lambda$ on four datasets.}
    \label{fig:hyperparameter}
\end{figure*}

The fourth group of baselines applies novel embedding 
(such as probabilistic embedding and box embedding) to recommendation tasks,
which improves the modeling and representation capabilities:
\begin{itemize}[leftmargin=*]
    \item \textbf{STOSA~\cite{DBLP:conf/www/FanLWWNZPY22/STOSA}:} This approach involves embedding items as stochastic Gaussian distributions and introducing a novel Wasserstein Self-Attention module to characterize item-item relationships.
    \item \textbf{CBox4CR~\cite{DBLP:conf/sigir/LiangZDXLY23/CBox4CR}:} It uses box embedding to represent items, so it can model user and item in a closed manner and introduce logical operators based on box embedding.
    \item \textbf{SRPLR~\cite{DBLP:conf/ijcai/YuanZX0LS023/SRPLR}:} It is a relatively novel framework that combines deep learning with symbolic learning, 
    which utilizes Beta embedding to model logical relationships. We set its backbone to SASRec.
\end{itemize}

\textbf{Implementation details.}
We implement PTSR using PyTorch and employ the Adam optimizer~\cite{DBLP:journals/corr/KingmaB14/Adam} with a learning rate of 0.0005,
and a weight decay coefficient searched within \{1e-9, 1e-8, 1e-7, 1e-6\}.
Regarding the number of pattern levels, we set it to 3 for the Yelp dataset
and 2 for the other three.
The margin $\gamma$ in Equation~\ref{eq:score} is set to 2.
Following previous studies~\cite{DBLP:conf/ijcai/YuanZX0LS023/SRPLR, DBLP:conf/sigir/LiangZDXLY23/CBox4CR}, the embedding size of either $\bm{\alpha}$ or $\bm{\beta}$ is fixed to 64.
The hyper-parameter $\lambda$ for enhancing the sequence-aware bias is discussed in Section \ref{section:hyper-parameter}.

For MSGIFSR, CBox4CR, STOSA, and SRPLR, we use the code released by the authors. 
For PMAN, since the code is not provided, we reproduce it based on the details in the original paper. Other methods are sourced from RecBole~\cite{DBLP:conf/cikm/ZhaoHPYZLZBTSCX22/RecBole}.
We employ a grid search strategy to meticulously explore optimal hyper-parameters for all baselines.
For fair comparisons, only implicit feedback interactions are available for all methods during training.
The maximum sequence length is set to 20 as the average sequence length is around 10 for all datasets. 
All methods are trained from scratch on a single NVIDIA GeForce RTX 2080 Ti GPU with a batch size of 512.

\subsection{Recommendation Performance Comparison (RQ1)}

We compare PTSR with other sequence models to validate the superiority in achieving state-of-the-art recommendation performance.
The experimental results have been presented in Table~\ref{tbl:performance-cmp},
from which we can draw two key conclusions:
\begin{itemize}[leftmargin=*]
  \item
  Traditional methods like Caser and GRU4Rec perform worse than attention-based methods, 
  suggesting that attention mechanisms are more suitable for modeling sequential data. 
  Although MSGIFSR introduces GRU and graph networks to model relationships within patterns and between patterns, respectively, 
  GRU fails to accurately capture pattern relationships (as evidenced by the performance of GRU4Rec). 
  Furthermore, MSGIFSR relies on repeated items in sequences to construct edges, 
  resulting in poor performance on datasets like Amazon, where repeated items are filtered out.
  Additionally, By applying novel (distribution-based and geometry-based) embeddings, 
  STOSA, CBox4CR, and SPRLR have demonstrated further performance improvements.
  However, the black-box nature of these models hinders users from understanding their decision-making process, thus affecting the customer experience.

  \item
  NAIS, RUM, and PMAN are three alternatives able to provide point-level analysis to justify an individual item's contribution to their recommendation results.
  However, it is clear that their interpretability comes at a huge cost in terms of performance, which is nearly 17.6\% worse compared to state-of-the-art sequential approaches.
  Specifically, while PMAN is based on self-attention, 
  it introduces a binary mask that indiscriminately filters attention weights for all users, 
  potentially discarding valuable information. 
  Additionally, by removing the feed-forward network, it loses part of its nonlinear modeling capability.
  In contrast, PTSR with Gamma embedding strikes a good balance between performance and interpretability. 
  It achieves better recommendation performance on the five datasets, with an average improvement of 5.23\%.

\end{itemize}

\begin{table*}[!t]
\centering
\setlength{\tabcolsep}{1.1mm}
\renewcommand\arraystretch{1.5}
\caption{
Performance of single-level and multi-level on four datasets. 
1, 2, etc. represent single-level, indicating the use of a single level; 
\{1, 2\}, etc. represent multi-level, indicating the simultaneous use of multiple levels.
The best results are marked in \textbf{bold}.
}
\label{tbl:level-cmp}
\begin{tabular}{cl|cccc|cccc|cccc|cccc}
\toprule
\multicolumn{2}{c|}{\multirow{2}{*}{Level}} & \multicolumn{4}{c|}{Beauty}                                            & \multicolumn{4}{c|}{Toys}                                              & \multicolumn{4}{c|}{Tools}                                             & \multicolumn{4}{c}{Yelp}                                              \\ 
\cmidrule(lr){3-6} \cmidrule(lr){7-10} \cmidrule(lr){11-14} \cmidrule(lr){15-18}
\multicolumn{2}{c|}{}                                  & N@5             & N@10            & HR@5            & HR@10           & N@5             & N@10            & HR@5            & HR@10           & N@5             & N@10            & HR@5            & HR@10           & N@5             & N@10            & HR@5            & HR@10           \\ \midrule
\multirow{3}{*}       & 1               & 0.3268          & 0.3621          & 0.4372          & 0.5462          & 0.3085          & 0.3434          & 0.4103          & 0.5185          & 0.2215          & 0.2579          & 0.3125          & 0.4252          & 0.4668          & 0.5184          & 0.6408          & 0.7997          \\
                                    & 2               & 0.3140          & 0.3492          & 0.4259          & 0.5350          & 0.3020          & 0.3387          & 0.4066          & 0.5205          & 0.2253          & 0.2626          & 0.3173          & 0.4327          & 0.4534          & 0.5063          & 0.6302          & 0.7927          \\
                                    & 3               & 0.3056          & 0.3405          & 0.4159          & 0.5235          & 0.2942          & 0.3312          & 0.4019          & 0.5161          & 0.2237          & 0.2604          & 0.3161          & 0.4299          & 0.4464          & 0.5006          & 0.6210          & 0.7877          \\ \midrule
\multirow{3}{*}      & \{1\}           & 0.3268          & 0.3621          & 0.4372          & 0.5462          & 0.3085          & 0.3434          & 0.4103          & 0.5185          & 0.2215          & 0.2579          & 0.3125          & 0.4252          & 0.4668          & 0.5184          & 0.6408          & 0.7997          \\
                                    & \{1,2\}         & \textbf{0.3404} & \textbf{0.3747} & \textbf{0.4512} & \textbf{0.5573} & \textbf{0.3276} & \textbf{0.3614} & \textbf{0.4302} & \textbf{0.5347} & \textbf{0.2458} & \textbf{0.2815} & \textbf{0.3377} & \textbf{0.4485} & 0.5004          & 0.5482          & \textbf{0.6665} & \textbf{0.8135} \\
                                    & \{1,2,3\}       & 0.3367          & 0.3701          & 0.4425          & 0.5454          & 0.3251          & 0.3571          & 0.4276          & 0.5266          & 0.2409          & 0.2749          & 0.3252          & 0.4309          & \textbf{0.5025} & \textbf{0.5498} & 0.6651          & 0.8107          \\ \bottomrule
\end{tabular}
\end{table*}

\begin{table*}[!t]
\centering
\setlength{\tabcolsep}{2.3mm}
\renewcommand\arraystretch{1.15}
\caption{
Key items recall (Recall@1,2,3,5) across three Amazon datasets.
Given a target item, the model needs to retrieve as many key items as possible
from the interaction sequence with specific relationships, 
including `Also-viewed', `Also-bought', and `Bought-together'.
}
\label{table-key-recall}
\begin{tabular}{l|l|cccc|cccc|cccc}
\toprule
\multicolumn{1}{l}{}             &        & \multicolumn{4}{c|}{Beauty}                                        & \multicolumn{4}{c|}{Toys}                                          & \multicolumn{4}{c}{Tools}                                         \\
\midrule
\multicolumn{1}{l|}{Relationship} & Method & R@1            & R@2            & R@3            & R@5            & R@1            & R@2            & R@3            & R@5            & R@1            & R@2            & R@3            & R@5            \\
\midrule
                                  & PTSR                          & \textbf{54.72}               & \textbf{67.26}               & \textbf{76.83}               & \textbf{87.81}               & \textbf{68.72}               & \textbf{75.63}               & \textbf{82.24}               & \textbf{91.46}               & \textbf{52.11}               & \textbf{65.25}               & \textbf{74.88}               & \textbf{87.21}               \\
                                  & RUM                           & 37.25                        & 46.99                        & 55.79                        & 73.75                        & 50.45                        & 55.32                        & 61.94                        & 79.80                        & 36.42                        & 44.75                        & 54.71                        & 75.41                        \\
                                  & NAIS                          & 19.15                        & 29.54                        & 42.43                        & 66.93                        & 24.21                        & 35.44                        & 49.46                        & 75.73                        & 34.61                        & 45.08                        & 58.02                        & 77.54                        \\
                                  & {PMAN} & {37.13} & {46.90} & {55.71} & {70.50} & {47.95} & {53.59} & {60.43} & {75.63} & {34.21} & {42.88} & {52.89} & {72.79} \\
                                  & {SASRec} & {29.83} & {42.13} & {55.43} & {76.66} & {45.34} & {56.72} & {67.81} & {84.52} & {30.58} & {44.24} & {61.65} & {80.49} \\
\multirow{-5}{*}{Also-viewed}     & {STOSA}  & {27.66} & {40.32} & {52.19} & {72.06} & {25.90} & {36.31} & {47.86} & {71.99} & {14.08} & {25.59} & {39.67} & {66.89} \\ \midrule
                                  & PTSR                          & \textbf{56.08}               & \textbf{68.30}               & \textbf{78.42}               & \textbf{89.63}               & \textbf{70.13}               & \textbf{76.51}               & \textbf{82.96}               & \textbf{92.07}               & \textbf{66.67}               & \textbf{74.65}               & \textbf{82.37}               & \textbf{92.73}               \\
                                  & RUM                           & 38.54                        & 48.24                        & 57.57                        & 75.94                        & 51.32                        & 56.06                        & 62.87                        & 80.59                        & 41.89                        & 49.22                        & 58.31                        & 78.65                        \\
                                  & NAIS                          & 21.59                        & 31.76                        & 44.55                        & 69.03                        & 26.39                        & 37.41                        & 51.52                        & 77.22                        & 40.35                        & 50.66                        & 63.13                        & 83.07                        \\
                                  & {PMAN} & {37.65} & {47.60} & {57.00} & {72.13} & {48.42} & {54.09} & {61.19} & {75.97} & {38.85} & {46.97} & {56.25} & {74.01} \\
                                  & {SASRec} & {30.57} & {42.62} & {55.85} & {76.66} & {45.34} & {56.85} & {68.00} & {84.58} & {38.61} & {50.01} & {62.56} & {81.82} \\
\multirow{-5}{*}{Also-bought}     & {STOSA}  & {29.13} & {41.05} & {52.41} & {72.97} & {27.87} & {38.23} & {49.21} & {72.53} & {20.77} & {32.78} & {45.73} & {70.76} \\ \midrule
                                  & PTSR                          & \textbf{53.41}               & \textbf{71.46}               & \textbf{83.12}               & \textbf{94.10}                & \textbf{64.16}               & \textbf{76.31}               & \textbf{86.66}               & \textbf{95.82}               & \textbf{69.67}               & \textbf{82.04}               & \textbf{90.06}               & \textbf{96.85}               \\
                                  & RUM                           & 32.63                        & 44.88                        & 54.37                        & 73.83                        & 45.13                        & 50.81                        & 60.68                        & 82.43                        & 42.89                        & 51.67                        & 61.98                        & 82.43                        \\
                                  & NAIS                          & 10.88                        & 19.32                        & 33.90                        & 65.56                        & 12.50                        & 24.93                        & 41.69                        & 76.78                        & 44.87                        & 57.12                        & 71.63                        & 89.02                        \\
                                  & {PMAN} & {35.03} & {47.05} & {56.54} & {72.68} & {43.03} & {49.00} & {58.88} & {77.78} & {40.59} & {49.57} & {59.87} & {78.61} \\
                                  & {SASRec} & {21.23} & {36.20} & {52.34} & {77.97} & {33.19} & {51.19} & {67.62} & {88.51} & {34.31} & {51.67} & {69.05} & {87.49} \\
\multirow{-5}{*}{Bought-together} & {STOSA}  & {23.18} & {39.32} & {54.78} & {77.83} & {15.43} & {28.16} & {44.06} & {73.69} & {17.78} & {31.71} & {46.13} & {73.73} \\ \bottomrule
\end{tabular}
\end{table*}

\subsection{Ablation Study (RQ2)} \label{sec:ablation}
The results of the ablation experiments are presented in Table~\ref{tbl:ablation}, 
which is divided into two main parts. 
The first part examines the effect of removing certain components on PTSR,
while the second part investigates the impact of combining different positional encodings with PTSR.
In the following, we describe the meaning of each part and analyze its effects:

\begin{figure*}[htbp]
    \centering
    \setlength{\abovecaptionskip}{0.1cm}
    \includegraphics[width=0.9\linewidth]{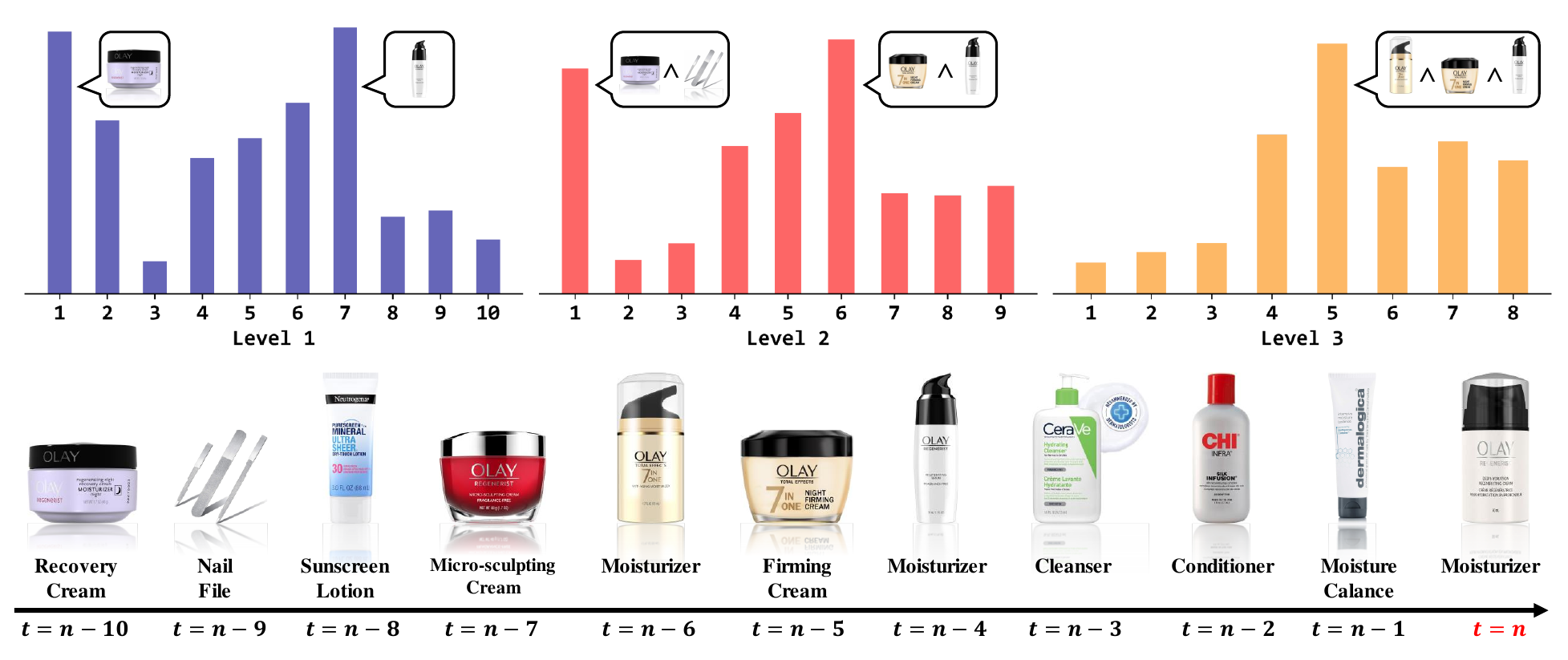}
    \caption{
    A case study of a real user.    
    The bottom panel presents the interaction history and the target item, 
    while the bar chart above displays the correlation distribution of patterns at three levels 
    (the higher the value, the stronger the correlation between the pattern and the target item, and vice versa).
    `$\land$' indicates a conjunction relationship.
    }  
    \label{fig:interpretable_case}
\end{figure*}

\begin{itemize}[leftmargin=*]
    \item \textit{w/o W (Weight)}: 
    After calculating the distance between the pattern and the candidate item, 
    we convert the negative distance into a weight 
    using a softmax function within the weight unit. 
    This weight is then multiplied by the distance to determine the pattern's contribution. 
    If the weight unit is removed, the distance itself is directly used as the pattern's contribution.
    Compared to \textit{DEFAULT}, this variant is significantly less effective. 
    The primary reason for this is that without the weights, 
    the distances of both the key and noise patterns converge around the margin, 
    leading to lower differentiation. 
    Adding the weight units emphasizes the decisive role of the key pattern in scoring 
    by assigning it a larger weight.
    \item \textit{w/o B (Bias)}:
    We utilize sequence-aware bias to enhance PTSR's ability to capture sequential information. 
    As illustrated in Table~\ref{tbl:ablation}, 
    the removal of this bias significantly affects the Beauty and Toys datasets, 
    while its impact on the Tools and Yelp datasets is relatively minor, 
    underscoring the critical role of this component for PTSR. 
    Notably, our model is capable of capturing certain sequential information 
    even without sequence-aware bias. 
    For instance, altering the sequence [$a$, $b$, $c$] to [$a$, $c$, $b$] 
    results in distinct patterns \{\{$a$, $b$\}, \{$b$, $c$\}\} and \{\{$a$, $c$\}, \{$c$, $b$\}\}
    with a window of size 2.
    Conversely, reversing the sequence to [$c$, $b$, $a$] yields the same patterns as [$a$, $b$, $c$].
    In such cases, bias further enhances the model's ability to perceive sequence changes.
    Therefore the order information is derived from both sequence-aware bias and inherent patterns,
    and a smaller influence of bias may suggest that the pattern itself plays a predominant role.
    \item \textit{w/o W+B (Weight and Bias)}:
    When both weight and bias are removed, 
    PTSR derives the score of the candidate item solely by aggregating the distances of patterns, 
    relying on the intrinsic characteristics of the patterns to perceive sequence information. 
    This approach is the most fundamental structure without any auxiliary mechanisms to aid optimization. 
    Unsurprisingly, this results in a significant performance decline of 15\% to 30\%, 
    unequivocally demonstrating the crucial combined impact of weight and bias on the efficacy of PTSR.
    \item \textit{Replace KL (KL-Divergence)}:
    Since PTSR is based on Beta/Gamma embedding, 
    we employ KL-Divergence to calculate the distance 
    between the distribution of the pattern and that of the target item. 
    To investigate the impact of KL-Divergence, 
    we replace it with cosine similarity as the metric. 
    Table~\ref{tbl:ablation} demonstrates a substantial decrease in PTSR's effectiveness. 
    We attribute this decline to several factors: 
    (1) the value range of cosine similarity is [0, 1], whereas KL-Divergence spans (0, +$\infty$), 
    allowing KL-Divergence to better capture differences; 
    (2) KL-Divergence is more adept at measuring distributional discrepancies; 
    and (3) KL-Divergence is asymmetric, which is particularly suitable for recommendation scenarios.
    For example, the probability of purchasing a computer followed by a desk 
    differs from the probability of purchasing a desk followed by a computer, 
    a distinction that KL-Divergence can effectively capture.
    \item \textit{Replace ProbE (Probabilistic Embedding)}:
    To explore the impact of embedding, we substitute probabilistic embedding with typical embedding. 
    Since the probabilistic operator and KL-Divergence are based on probability distributions, 
    we replace them with standard weighted sum and cosine similarity, respectively.
    Except for the Toys dataset, the experimental results show an even greater decrease in performance 
    compared to the replacement with KL-Divergence. 
    Thus, we argue that the efficacy of PTSR primarily 
    stems from the sophistication of probabilistic embedding, 
    which offers interpretable probabilistic operators. 
    Additionally, KL-Divergence further enhances the advantages of probabilistic embedding, 
    making PTSR significantly more effective than the baselines.
\end{itemize}

\subsection{Analysis of Pattern Level (RQ3)}
In Section~\ref{sec:metho}, we elucidate the roles of point-level and union-level. 
In this section, we will delve into a thorough examination of 
the performance of single-level and multi-level.
It should be noted that `point-level' and `union-level' refer to the granularity of items, 
with `point' denoting a single item and `union' indicating a combination of items. 
Similarly, `single-level' and `multi-level' refer to the granularity of levels, 
where `single' denotes a single level and `multi' signifies a combination of levels.

As shown in Table~\ref{tbl:level-cmp},
we conduct two sets of experiments. 
The first set is single-level, 
where we use a single sliding window to extract patterns from a specific level.
The second set is multi-level, where \{1,2,3\} means that we simultaneously use window sizes of 1, 2, and 3 for extraction.
The results reveal two observations:
\begin{itemize}[leftmargin=*]
    \item When only using one level of patterns (i.e., single-level), high levels do not exhibit the same performance as low levels. 
    This phenomenon indicates low levels, especially point-level, play a foundation role in sequential recommendation.

    \item When using multiple levels of patterns, the recommendation performance could be improved.
    Specifically, the best result is achieved with the combinations {1, 2} for Beauty, Toys, and Tools datasets, and {1, 2, 3} for the Yelp dataset. 
    This is attributed to the effectiveness of integrating different levels, as each captures user interests at varying granularities and offers distinct interpretability. 
\end{itemize}

\subsection{Impact of Hyperparameter \texorpdfstring{$\lambda$}{lambda} (RQ4)} \label{section:hyper-parameter}
We investigate the sensitivity of the hyperparameter $\lambda$ introduced for controlling the effect of sequence-aware bias. 
We evaluate it on four datasets with values ranging from 0 to 1. 
As illustrated in Fig.~\ref{fig:hyperparameter}, 
for most datasets (Beauty, Toys, and Yelp), 
the performance initially increases and reaches its peak at approximately $\lambda=0.4$. 
However, further increasing this value deteriorates the performance as the distance-based weight becomes overly weakened. 
Despite this, PTSR with $\lambda \approx 1$ consistently outperforms other recommendation methods (SRPLR, BERT4Rec, and CBox4CR), 
reaffirming the superiority of PTSR. 
An exception is the Tools dataset, in which the explicit sequence-aware bias has a negligible effect. 
We hypothesize that the pattern extraction process itself retains sufficient sequence information, which appears to be adequate for the Tools dataset.

\subsection{Interpretability Analysis (RQ5)}
\label{sec:interpretability}
In this section, we analyze and compare the interpretability of PTSR from both specific examples and statistical perspectives.

\begin{figure}[!t]
    \centering
    \setlength{\abovecaptionskip}{0.cm}
    \includegraphics[width=0.9\linewidth]{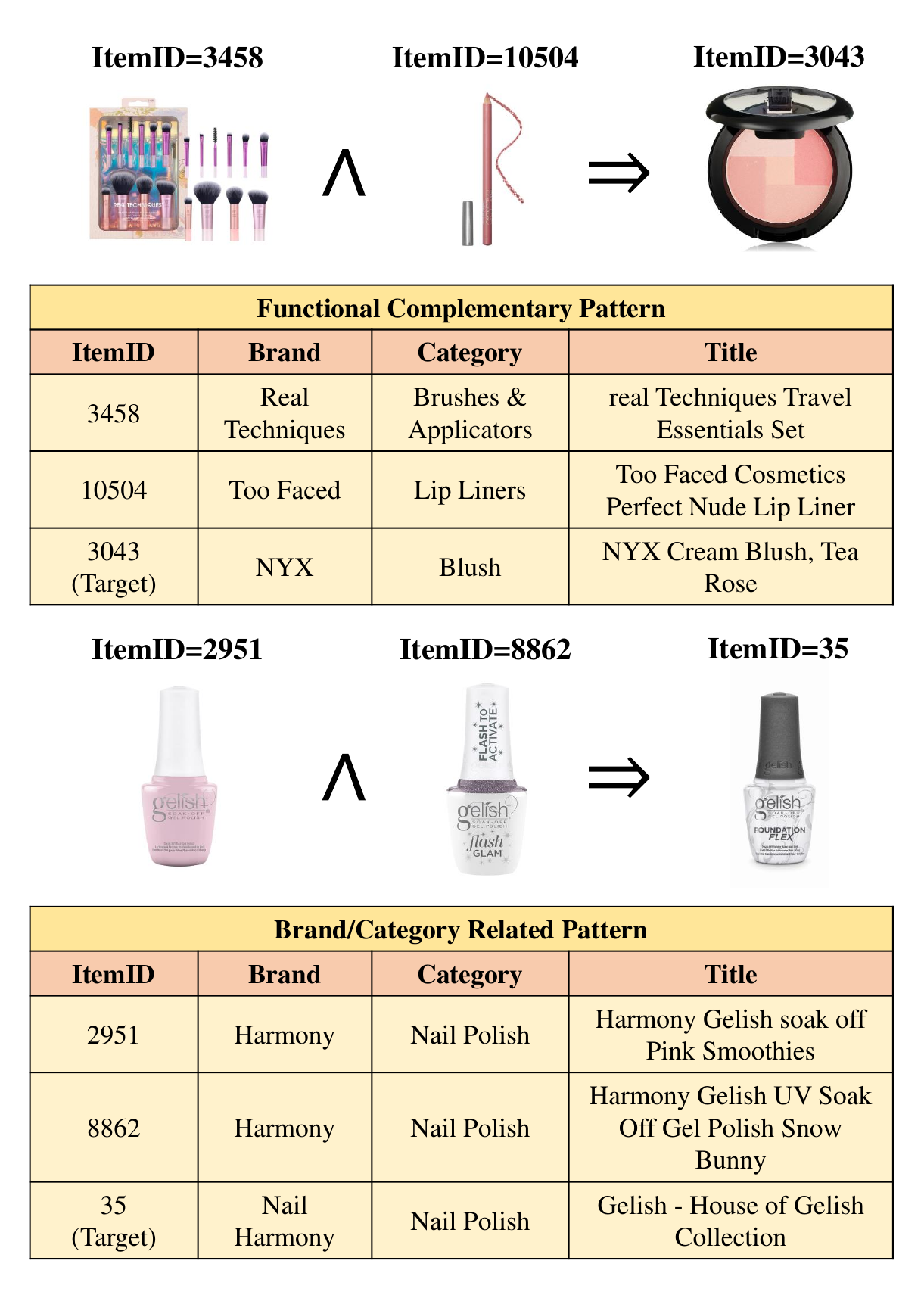}
    \caption{
        Showcase of different types of union-level patterns.
    }
    \label{fig:union_case}
\end{figure}

\textbf{Case Analysis.}
As shown in Fig.~\ref{fig:interpretable_case}, 
we visualize the distribution of pattern importance from different levels in PTSR recommendations. 
Level 1, or `point-level', focuses on individual items, 
while levels 2 and above, or `union-level', emphasize the combined effects of items. 
It can be observed that there are both commonalities and differences between different levels. 
(1) \textbf{Commonalities}: For the target item, an `OLAY' moisturizer, 
there is a strong correlation with other `OLAY' products in sequence, reflecting user purchasing habits. 
Our model shows strong correlations across all three levels, 
with multi-level patterns collaborating to enhance the influence of key items and capture user interests.
(2) \textbf{Differences}: In level 1 and 2, 
`Recovery cream' and `Nail File' show high correlations, mostly because they often appear together with the target item.
However, in level 3, adding `Sunscreen Lotion' significantly lowers the correlation of the pattern with these two products, 
indicating they do not form an effective pattern from the PTSR's perspective. 
Conversely, some products on the right side of the sequence, 
while weakly correlated in level 1 and 2, show strong correlations in level~3, 
demonstrating the complementary nature of the multi-level patterns in identifying effective patterns.
Additionally, Fig.~\ref{fig:union_case} illustrates various types of patterns captured by PTSR in real-world scenarios. 
Remarkably, even without providing any information beyond itemID, 
PTSR can still identify meaningful combinations of items through logical modeling.

\textbf{Statistical Analysis.}
Following~\cite{DBLP:conf/sigir/McAuleyTSH15/AmazonDataset, DBLP:conf/wsdm/BingZD23/CogER}, 
items that have the `Also-viewed', `Also-bought' or `Bought-together'
relation with the target item are considered important.
Therefore, the desired model should assign a high correlation/contribution value to these key items.
We use Recall@K to evaluate the recall effectiveness of key items. 
Specifically, each model assigns weights to items in sequence, 
with some items having relationships with the target. 
We rank items in descending order based on their weights and check whether the key items are recalled by the model when different values of K are set.

Table~\ref{table-key-recall} reports the comparisons. 
We compare five different types of models, 
including those with transparent structures (such as RUM, NAIS, and PMAN) 
as well as classic attention-weighted models (SASRec and STOSA).
When K is 5, these methods perform well; 
however, at K = 1, their performance declines, highlighting limitations in accurately identifying key items. 
This may stem from the models' inherent weaknesses 
or their inability to adjust attention weights dynamically based on target changes. 
Notably, PTSR consistently outperforms others across all datasets and relationship types, 
enhancing both recommendation performance and model interpretability.

\begin{figure}[!t]
    \centering
    \setlength{\abovecaptionskip}{0.2cm}
    \includegraphics[width=0.98\linewidth]{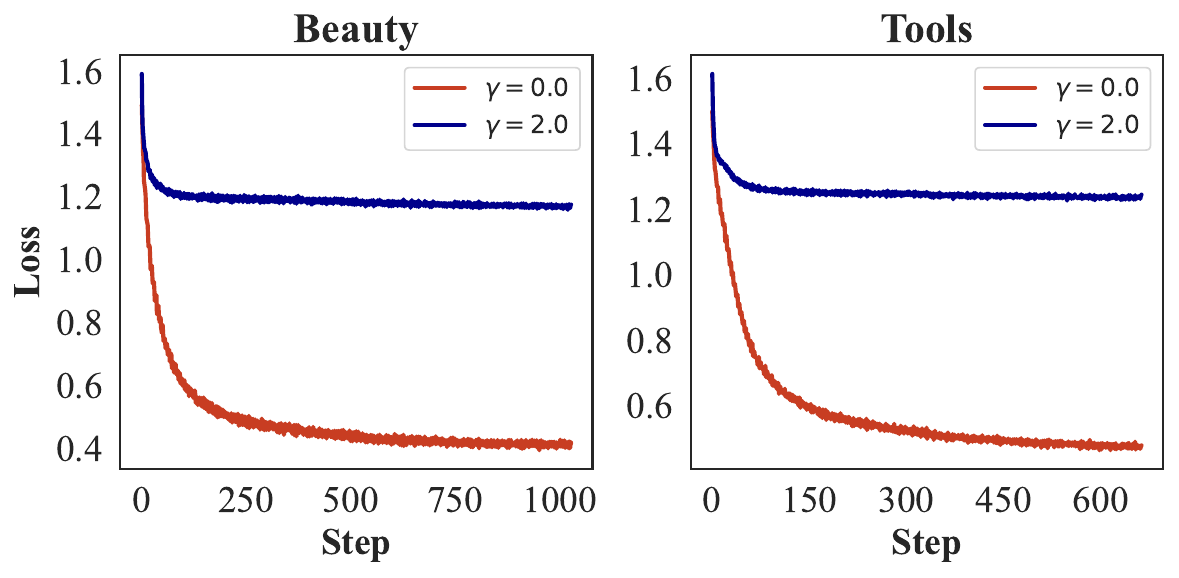}
    \caption{
        Training losses of PTSR on Beauty and Tools dataset.
    }
    \label{fig:training_loss}
\end{figure}

\subsection{Training Stability Analysis (RQ 6)}
In PTSR, unlike traditional optimization of inner products, 
the model primarily optimizes the KL-Divergence between different probability distributions. 
The training stability of the model is a critical issue. 
As shown in Fig.~\ref{fig:training_loss}, 
when we adjust the decision boundary \(\gamma\) in Equation~\eqref{eq:score} to an appropriate value (set to 2 in experiments), 
the training loss on different datasets can steadily decrease and be optimized to a lower value. 
This indicates that with the support of the weight and bias components, 
probability representations can be properly optimized, 
and KL-Divergence is a reliable optimization metric.

\section{Conclusions and Future Work}\label{sec:clu}
In this work, we design a probabilistic embedding-based transparent framework 
for interpretable sequential recommendation. 
Our primary motivation is that items in an interaction sequence influence the recommendation of the next item, 
either individually (low-order, point-level) or in combination (high-order, union-level). 
We therefore propose using sliding windows to extract multi-level patterns 
and employing a probabilistic embedding-based conjunction operator to model high-order patterns, 
ensuring complete transparency of PTSR. 
The weight part is designed to highlight the different importance of patterns 
and aid optimization. 
Sequence-aware bias aims to help the model perceive sequential information.
Experimental results demonstrate PTSR is not only highly transparent and interpretable 
but also significantly outperforms classic sequential recommendation models 
w.r.t. recommendation performance.

Despite the promising results achieved by PTSR, 
several limitations remain. 
First, items that are spaced apart in a sequence can also form valid patterns, 
but our current union-level approach can only model consecutive items. 
Finding a way to dynamically select items from the sequence to form patterns is an interesting direction. 
Second, enhancing the model's logical reasoning capabilities requires incorporating different operators, 
particularly the negation operator, to form disjunctive or conjunctive normal forms,
as used in~\cite{DBLP:conf/wsdm/JiLXXTGWZ23/CCR, DBLP:conf/www/ChenSLZ21/NCR}.
However, we currently only use conjunction operations, 
and exploring how to introduce negation operators to model users' explicit negative feedback behaviors is worth investigating.
Third, we use softmax to obtain weights for different patterns, 
but as the sequence length increases, 
the differences between weights may become less distinguishable, 
potentially leading to poor performance on longer sequences. 
We will gradually address these issues in our future work.

\ifCLASSOPTIONcaptionsoff
  \newpage
\fi

\bibliographystyle{IEEEtran}
\bibliography{TKDE}

\begin{IEEEbiography}
  [{\includegraphics[width=1in,height=1.25in,keepaspectratio]{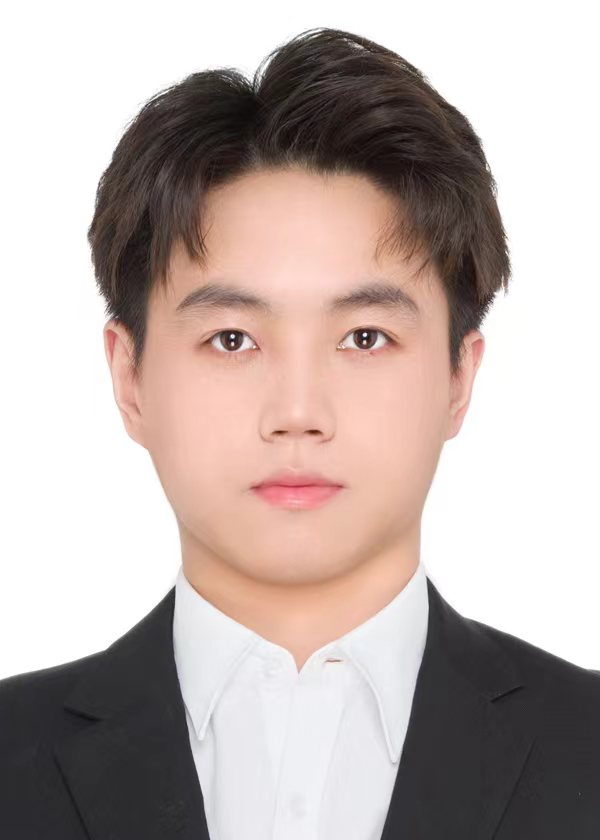}}]
  {Kun Ma} is currently pursuing the master’s degree with the Department of Computer Science and Technology, East China Normal University, China. His research interests include machine learning, data mining, and explainable recommender system.
\end{IEEEbiography}

\begin{IEEEbiography}
  [{\includegraphics[width=1in,height=1.25in,clip,keepaspectratio]{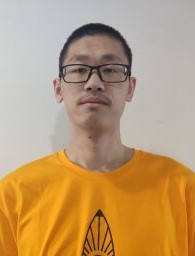}}]
  {Cong Xu}
  is currently pursuing the Ph.D. degree 
  in the School of Computer Science and Technology,
  East China Normal University, Shanghai, China.
  His main research interests include machine learning and data mining.
\end{IEEEbiography}


\begin{IEEEbiography}
  [{\includegraphics[width=1in,height=1.25in,keepaspectratio]{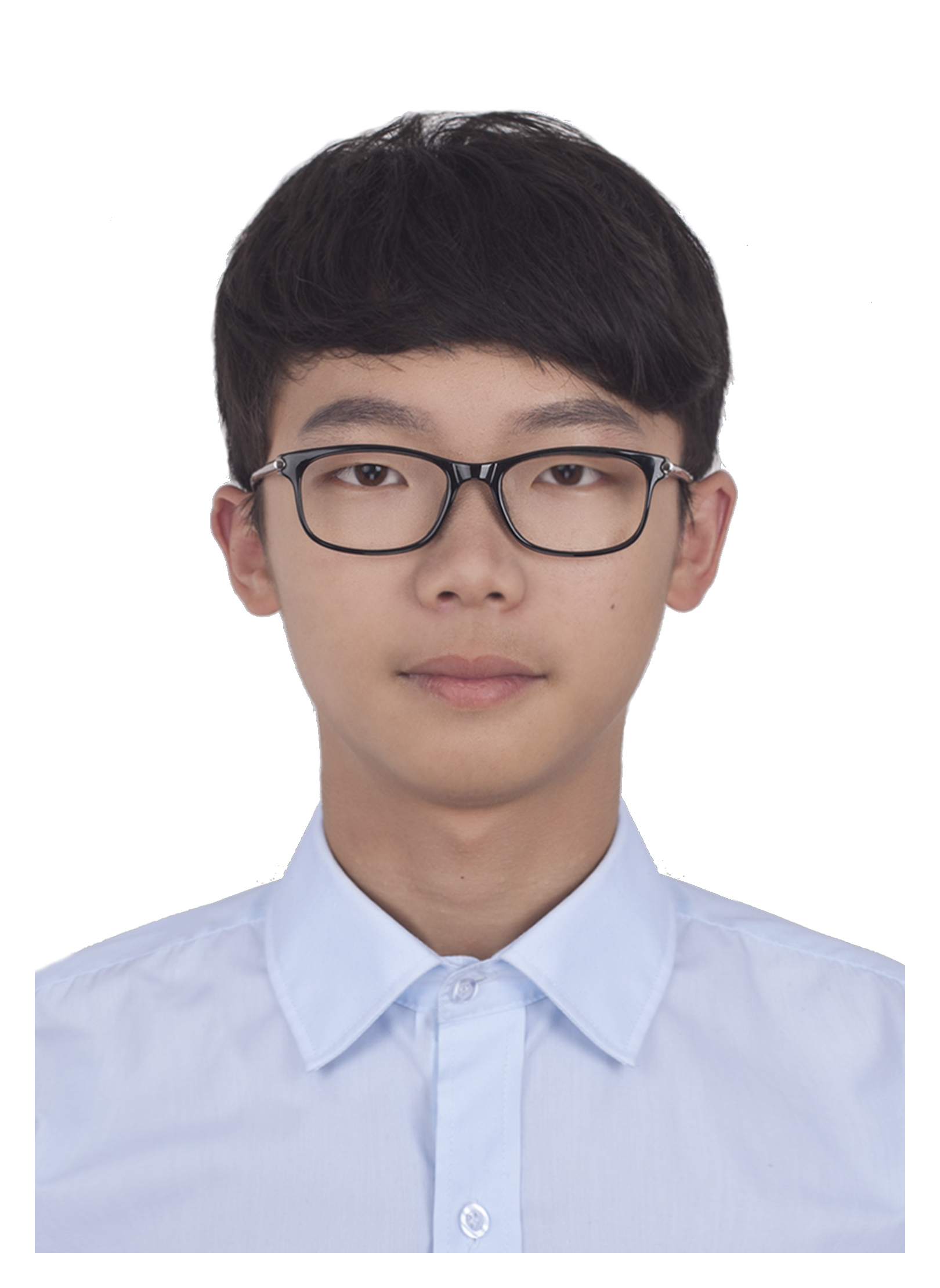}}]
  {Zeyuan Chen}
    received the master degree from East China Normal University, China, in 2022. His research interests include recommender systems and data mining.
\end{IEEEbiography}

\vspace{-15cm}

\begin{IEEEbiography}[{\includegraphics[width=1in,height=1.25in,clip,keepaspectratio]{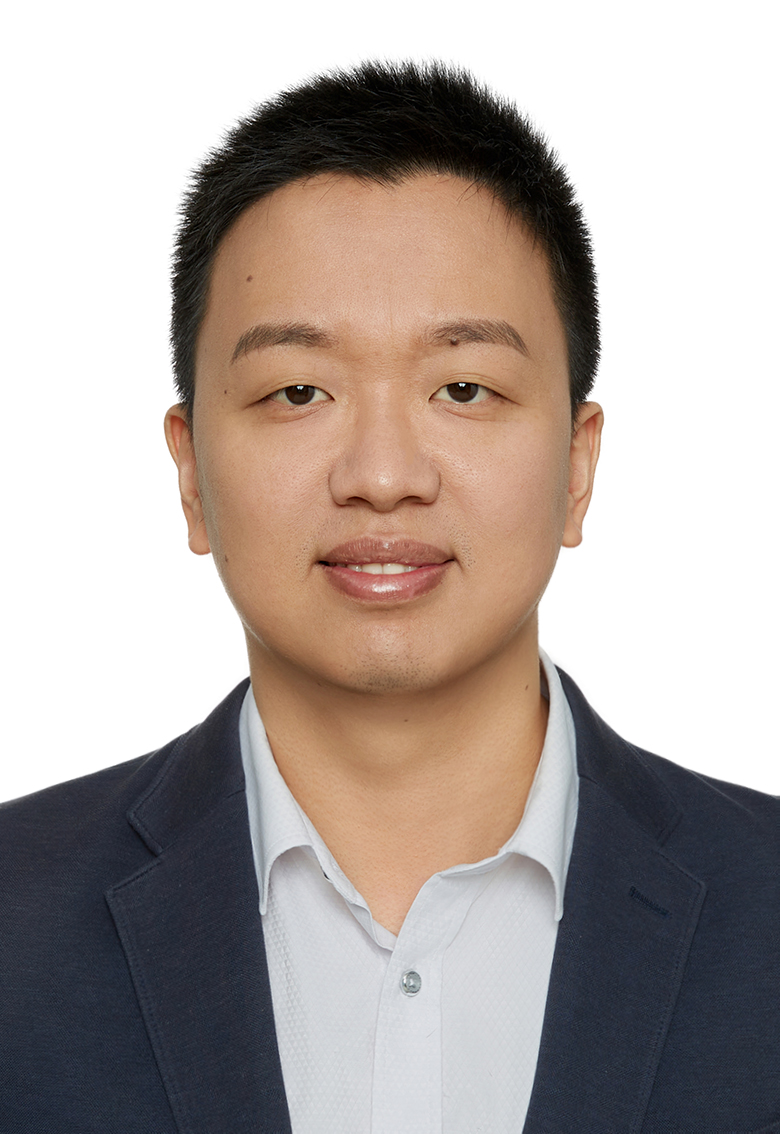}}]{Wei Zhang} received his Ph.D. degree in computer science and technology from Tsinghua university, Beijing, China, in 2016. He is currently a professor in the School of Computer Science and Technology, East China Normal University, Shanghai, China.
His research interests span data mining and machine learning.
his current interests focus on deep recommender systems, interpretable machine learning, large language models, and AI for education.
He is a senior member of China Computer Federation.
\end{IEEEbiography}

\end{document}